\documentclass[]{emulateapj}

\usepackage{graphics,graphicx,xspace,natbib,amssymb,xcolor}
\usepackage[caption=false]{subfig}
\usepackage{amsmath}

\newcommand{\psbs}{post-starburst galaxies\xspace}
\newcommand{\sersic}{S\'{e}rsic\xspace}
\newcommand{\sig}{$\Sigma_1$\xspace}
\newcommand{\sige}{$\Sigma_{\rm{e}}$\xspace}
\newcommand{\av}{$A_{\rm{v}}$\xspace}
\usepackage[utf8]{inputenc}

\shorttitle{Dissecting Structural Relations}
\shortauthors{Suess et al.}

\begin{document}

\title{Dissecting the size-mass and \sig-mass relations at $1.0<\lowercase{z}<2.5$:\\galaxy mass profiles and color gradients as a function of spectral shape}
\author{Katherine A. Suess\altaffilmark{1}, Mariska Kriek\altaffilmark{1}, Sedona H. Price\altaffilmark{2}, Guillermo Barro\altaffilmark{3}} 

\submitted{Submitted to ApJ}

\altaffiltext{1}{Astronomy Department, University of California, Berkeley, CA 94720, USA}
\altaffiltext{2}{Max-Planck-Institut f{\"u}r extraterrestrische Physik, Postfach 1312, Garching, 85741, Germany}
\altaffiltext{3}{Department of Physics, University of the Pacific, 3601 Pacific Ave, Stockton, CA 95211, USA}
\email{suess@berkeley.edu}

\begin{abstract}
We study how half-mass radii, central mass densities (\sig), and color gradients change as galaxies evolve. We {separate} $\sim7,000$ galaxies into sixteen groups with similar spectral shapes; each group represents a different evolutionary stage. We find that different galaxy types populate different regions of both size-mass and \sig-mass space. The nine star-forming groups lie along the integrated star-forming \sig-mass relation. However, these star-forming groups form steep parallel relations in the size-mass plane, with slopes similar to the quiescent size-mass relation. These steep slopes can be explained as a transformation of the star-forming \sig-mass relation and its scatter. We identify three types of transitional galaxies. Green valley and post-starburst galaxies are similarly compact at $z>1.5$; however, their distinct color gradients indicate that the two populations represent different pathways to quenching. Post-starburst galaxies have flat color gradients and compact structures, consistent with a fast quenching pathway which requires structural change and operates primarily at high redshift. Green valley galaxies have negative color gradients, and are both larger and more numerous towards lower redshift. These galaxies are consistent with slow quenching without significant structural change. We find that dusty star-forming galaxies at $z\gtrsim2$ are very compact, and may represent the ``burst" before post-starburst galaxies; at $z\lesssim2$, dusty star-forming galaxies are extended and have shallow color gradients consistent with slow quenching. Our results suggest that star-forming galaxies grow gradually up the \sig-mass relation until (a) they naturally reach the high \sig values required for quiescence, or (b) a compaction-type event rapidly increases their \sig.
\end{abstract}


\section{Introduction}

The structures of galaxies change as they grow in mass and move through different evolutionary phases. Previous studies have found that star-forming disks grow ``inside-out," gradually increasing both their sizes and their stellar masses as they form stars \citep[e.g.,][]{wuyts11,wuyts13,vanderwel14,abramson14,nelson16}. Even after galaxies cease forming stars, this inside-out growth continues: the sizes of quiescent galaxies increase over cosmic time due to minor mergers \citep[e.g.,][]{bezanson09,naab09,hopkins09,newman12,vandesande13,suess19b}. 

However, the largest structural changes appear to be closely linked to the still-mysterious ``quenching" process that shuts down star formation in galaxies. 
In addition to characteristic differences in their star formation rates, colors, and stellar masses \citep[e.g.,][]{blanton03,kauffmann03,shen03,noeske07,wuyts11}, average star-forming and quiescent galaxies have significantly different structures.
Star-forming galaxies tend to have disk-like light profiles, whereas quiescent galaxies have more significant bulge components and higher \sersic indices \citep[e.g.,][]{bell08,wuyts11,vandokkum11,lang14,bluck14}. 
Star-forming and quiescent galaxies also have distinct sizes: quiescent galaxies are smaller at fixed mass than their star-forming counterparts, and follow a steeper relation in size-mass space \citep[e.g.,][]{shen03,vanderwel14,mowla18,suess19a}. 
Understanding why and how these structural changes take place could be key to understanding the physical mechanism(s) responsible for shutting down star formation and creating the galaxy bimodality.

Despite characteristic differences in the sizes and masses of star-forming and quiescent galaxies, neither size nor mass alone is a clear and unique predictor of quiescence \citep[e.g.,][]{omand14,whitaker17}. 
Combining these quantities to estimate the 
effective surface mass density, \sige, 
can effectively separate star-forming and quiescent galaxies both in the local universe \citep[e.g.,][]{kauffmann03,brinchmann04} and beyond \citep[e.g.,][]{franx08,maier09}. 
Recent work has suggested that the central mass density in a {\it fixed} aperture of one kiloparsec, \sig, provides an even clearer predictor of quiescence \citep[e.g.,][]{cheung12,fang13,vandokkum14,barro17,whitaker17,lee18,woo19}. \sig is also easier to interpret than \sige: \sige can increase or decrease as a galaxy evolves, depending on the evolution of both the galaxy's mass density and its effective radius. In contrast, \sig does not directly depend on the evolution of the effective radius, and can only decrease with significant mass loss or adiabatic expansion  \citep[e.g.,][]{damjanov09,poggianti13,barro17}.
In the \sig-mass plane, both star-forming and quiescent galaxies lie on tight and well-defined relations with similar slopes of $\sim0.8-1$, with quiescent galaxies offset to higher \sig than star-forming galaxies \citep[e.g.,][]{fang13,barro17}. 
These measurements seem to indicate that building a dense core and reaching high \sig is a necessary prerequisite for quenching \citep[e.g.,][]{fang13,barro17,vandokkum15,tacchella15_apj,mosleh17,whitaker17,suess20}.

Understanding the details of how this structural transformation takes place requires investigating the sizes and \sig values of possible quiescent progenitors.
Several types of these transitional galaxies have been identified in the literature.
Post-starburst galaxies, which recently and rapidly stopped a major star-forming epoch, have small sizes and high \sig \citep[e.g.,][]{whitaker12_psb,yano16,almaini17,belli19,suess20}. These galaxies are much more compact than average star-forming galaxies, indicating that the process that shuts down star formation in post-starburst galaxies also alters their structures. Green valley galaxies, which have intermediate colors and specific star formation rates,
seem to have relatively large sizes indicating that their star formation may cease without a need for sudden structural transformation \citep[e.g.,][]{martin07,mendez11,wu18}. Finally, massive compact star-forming galaxies have similar structures and number density evolution as compact quiescent galaxies at $z\sim2$, suggesting an evolutionary link between the two populations \citep[e.g.,][]{barro13,barro14,vandokkum15}. 
The varying structures of these different classes of possible quiescent progenitors have led to an emerging picture of at least two distinct pathways to quiescence, a ``fast" quenching process which produces compact post-starburst galaxies, and a ``slow" quenching process that produces more extended green valley galaxies (e.g., \citealt{barro13,woo15,belli19,woo19}; though see also \citealt{lilly16,abramson16}).

The majority of these detailed structural studies at $z>0$ have analyzed the light profiles of quiescent galaxies and one or two classes of transitional galaxies. This approach has two key setbacks. 
First, pre-selecting just a few types of possible quiescent progenitors means that these studies may have used incomplete or biased samples of transitional galaxies. 
Excluding the bulk of star-forming galaxies also prohibits building an understanding of how galaxies grow {before} they shut down their star formation. 
To build a complete picture of the way that galaxies grow and quench, we need to study the structural properties of a complete sample of galaxies across all phases of their evolution.

Second, most previous studies beyond the local universe have measured structural properties from galaxy light profiles. 
This approach neglects the effects of radial color gradients caused by variations in the underlying stellar populations of galaxies. Color gradients are known to exist in galaxies across a wide redshift range \citep[e.g.,][]{tortora10,szomoru12,szomoru13,lang14,mosleh17,suess19a,suess19b,mosleh20}, and the strength of color gradients depends on other galaxy properties such as stellar mass \citep[e.g.,][]{tortora10,suess19a,suess19b} and age along the quiescent sequence \citep[e.g.,][]{suess20}. 
These color gradients cause the light profiles of galaxies to deviate from their underlying mass profiles. Therefore, neglecting color gradients introduces a bias into structural studies and may obfuscate the structural changes that take place as galaxies transition from star-forming to quiescent. 

The next step forward, then, is to build an understanding of how the mass profiles of galaxies change as they grow and evolve over cosmic time--- how do galaxies move through the size-mass and \sig-mass planes? 
``Dissecting" these structural relations requires both a large sample of galaxies with mass profile measurements and a method to separate those galaxies into different evolutionary stages. 
This is now possible due to the \citet{suess19a} public catalog of mass profiles and half-mass radii for $\sim7,000$ galaxies at $1.0<z<2.5$ in the CANDELS fields. The wealth of ancillary data available in these well-studied extragalactic fields gives us strong constraints on the rest-frame spectral energy distributions (SEDs) of these galaxies. SEDs encode information about specific star formation rate (sSFR), dust, and age, providing a clear window into the evolutionary stage of each galaxy. Previous studies have shown that clustering galaxies based on the shapes of their rest-frame SEDs is a nearly model-independent way to select galaxies that are in similar evolutionary stages \citep[e.g.,][]{kriek11,yano16,forrest18}. By using this technique to group the galaxies in the \citet{suess19a} sample, we can study where a wide variety of galaxy types lie in the size-mass and \sig-mass planes.

In this paper, we present the structural properties of sixteen different groups of galaxies with similar rest-frame SED shapes. Using these groups, we show how the size-mass and \sig-mass relations change as a function of sSFR and redshift. Using this unique data, we suggest a new way to view the star-forming size-mass relation and present evidence that there are two distinct pathways to quench star formation in galaxies. Finally, we compare our results to various proposed galaxy evolution scenarios and suggest possible routes for galaxies to move through \sig-mass and size-mass space as they evolve.

This paper is organized as follows. In Section~\ref{sec:sampleselection}, we describe our sample of galaxies and summarize the methods used to calculate mass profiles and half-mass radii. In Section~\ref{sec:sed_groups}, we describe the methods used to separate our sample into groups with similar rest-frame SEDs. In Section~\ref{sec:groupResults}, we show where different groups of galaxies lie in size-mass and \sig-mass space. Section~\ref{sec:discuss_sf} discusses implications for the star-forming sequence, and Section~\ref{sec:discuss_qui} discusses implications for transitioning and quiescent galaxies. Finally, Section~\ref{sec:discuss_wrapup} places our results into a unified picture of how galaxies grow and quench. Throughout this paper we assume a standard $\Lambda$CDM cosmology with $\Omega_{\rm m}=0.3$, $\Omega_\Lambda=0.7$, and $h=0.7$. We also assume a \citet{chabrier03} initial mass function.

\section{Sample, Mass Profiles, \& Half-Mass Radii}
\label{sec:sampleselection}
For this work, we use the galaxy sample presented in \citet{suess19a}. This sample consists of all galaxies in the ZFOURGE survey \citep{straatman16} with high signal-to-noise ratio in the ZFOURGE detection band ($S/N_K > 10$), a ZFOURGE use flag equal to one, a match in both the 3D-HST photometric catalog and the \citet{vanderwel12} structural catalog, overlapping CANDELS imaging, $1.0 \le z \le 2.5$, and $\log{M_*/M_\odot} > 9.0$. The catalog of half-mass radii for this sample of galaxies is publicly available, and was released with \citet{suess19a}. The half-light radii and \sersic indices used in this work are taken from the \citet{vanderwel12} catalogs. To facilitate comparisons with other studies, we correct these half-light radii to rest-frame 5,000\AA\ using the procedure described by \citet{vanderwel14}. Both the half-light and half-mass radii used in this paper are measurements of the galaxy's major axis, not circularized radii. 

We use the photometric redshifts and stellar masses from the ZFOURGE catalog; the stellar masses are corrected to agree with the \citet{vanderwel12} structural measurements by multiplying the masses by the ratio of the F160W flux reported in the \citet{vanderwel12} catalog and the ZFOURGE catalog \citep{straatman16}. {We note that stellar mass estimates from SED fitting are affected by the choice of star formation history \citep[e.g.,][]{wuyts09}. Models which use ``non-parametric" star formation histories may more accurately recover stellar masses than models which use delayed exponential star formation histories \citep[e.g.,][]{iyer17,iyer19,leja19a,leja19b,lower20}. We use ZFOURGE stellar masses, calculated using delayed exponential star formation histories, for consistency with previous studies of the size-mass and \sig-mass relations as well as our half-mass radius measurement techniques \citep{suess19a}. The difference between these stellar masses and those estimated using non-parametric star formation histories on the order of 0.1 - 0.2 dex, and does not significantly depend on sSFR; these stellar mass uncertainties do not change the major conclusions of this paper.
}

The color gradients and half-mass radii used in this paper are calculated using the primary method described in detail in \citet{suess19a}. In summary, this method uses spatially-resolved SED fitting to calculate mass profiles.   
We use the high-resolution multi-band imaging and integrated photometry in the COSMOS, GOODS-S, and UDS fields obtained by the CANDELS program \citep{grogin11,koekemoer11} and PSF-matched by the 3D-HST team \citep{skelton14,momcheva16}. For each available filter, we measure the galaxy's flux in concentric elliptical annuli that follow the geometry of the best-fit structural parameters from the \citet{vanderwel12} catalog. Then, we use FAST \citep{kriek09} to fit these spatially-resolved SEDs with stellar population synthesis models to obtain the mass in each annulus. {These spatially-resolved FAST fits assume a \citet{chabrier03} IMF, the \citet{bc03} stellar population models, the (fixed-slope) \citet{kriek13} dust law, a delayed exponential star formation history, and solar metallicity.} This procedure yields the observed-space $M/L$ gradient profile. We then correct for the effects of the telescope point spread function (PSF) by using a forward modeling approach to convert this observed $M/L$ to an intrinsic $M/L$. We assume that the intrinsic $M/L$ is a power-law function of radius, then find the intrinsic $M/L$ profile that, when convolved with the PSF, best reproduces our observed $M/L$ profile. We multiply this best-fit intrinsic $M/L$ model by the galaxy's intrinsic light model to obtain the mass profile, then integrate to find the half-mass radius.  

\citet{suess19a} demonstrates that the half-mass radii measured in this way are robust, and are not significantly biased by the galaxy's stellar mass, redshift, or half-light radius. Furthermore, the half-mass radii determined using this method agree well with those obtained using other common methods for measuring half-mass radii \citep{lang14,chan16,szomoru10,szomoru12,szomoru13,mosleh20}. In addition to calculating half-mass radii, we use the mass profiles described above to calculate \sig, the surface mass density in each galaxy's central kiloparsec \citep[e.g.,][]{fang13}. Finally, we probe the strength of color gradients in each galaxy as the ratio of half-mass to half-light radii.

\section{Methods: Creating Composite SED Groups}
\label{sec:sed_groups}

\begin{figure}
    \centering
    \includegraphics[width=.5\textwidth]{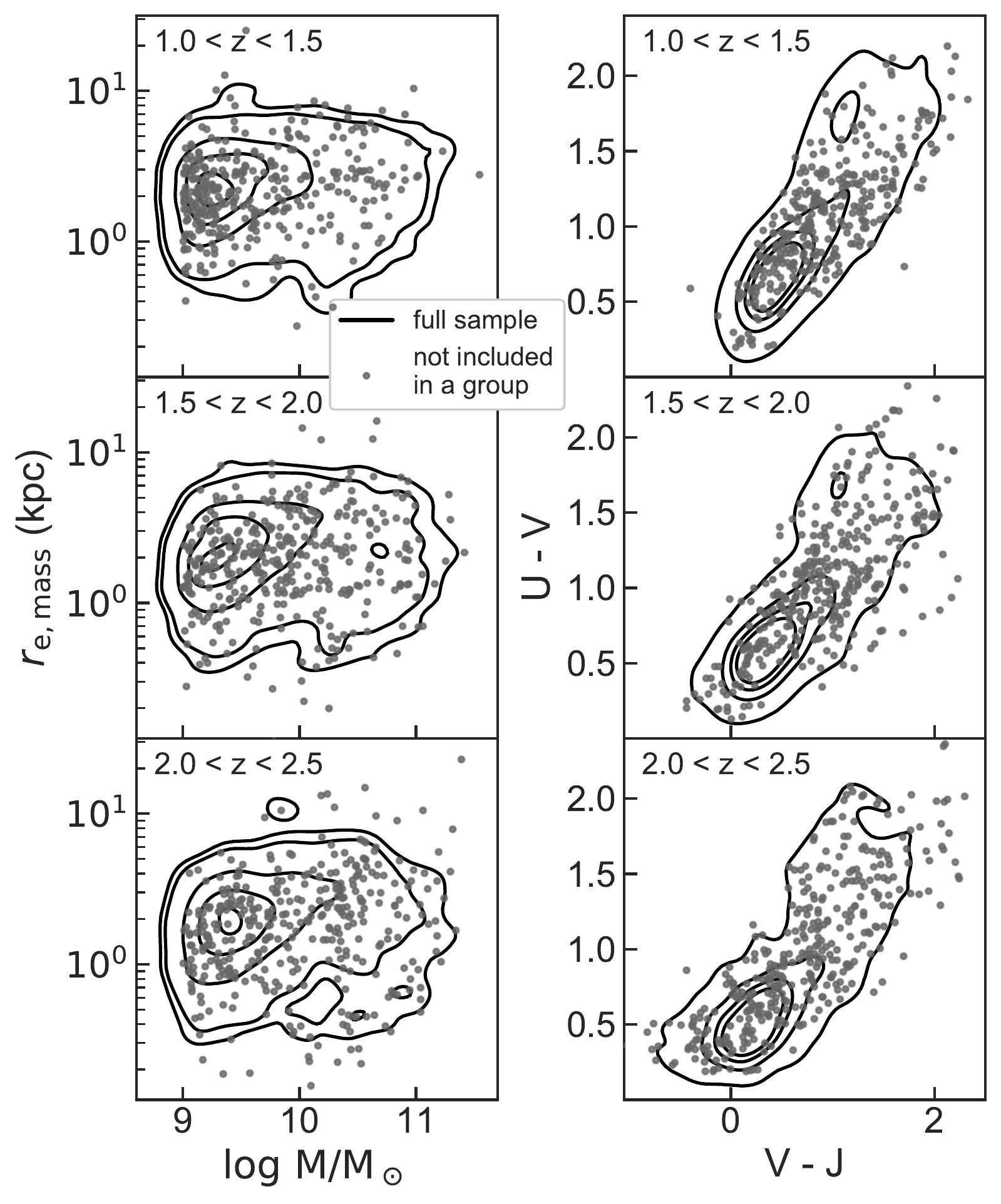}
    \caption{
    Size-mass and $UVJ$ distribution of our full sample (black contours) and galaxies that are not classified in a group (grey points). Each row shows a different redshift interval. The galaxies not included in a group are spread uniformly across size-mass and $UVJ$ space.} 
    \label{fig:ingroup}
\end{figure}

In this paper, we go beyond separating galaxies into just two groups--- star-forming or quiescent--- and ``dissect" the traditional size-mass and \sig-mass relations for many different types of galaxies.
We use the \citet{kriek11} technique to separate the $\sim$7,000 galaxies in our sample into groups with similar rest-frame SED shapes. The medium-band filters included as part of the ZFOURGE survey provide high-resolution SEDs, which are essential to create clean groups with this technique \citep{kriek11}. 
While this rest-frame SED grouping technique may appear similar to grouping by specific star formation rate (sSFR), it has several advantages. First, SED shape encodes multiple properties: in addition to sSFR, the SED shape is also influenced by dust, age, and star-formation timescale. Therefore, grouping by SED shape allows us to separate galaxies with the same sSFR but differences in other properties. Second, SED shape is a direct observable--- it is simply the galaxy's observed photometry--- while sSFR must be estimated by fitting models to the observed photometry. The method described below relies on models only to estimate photometric redshifts and to interpolate rest-frame colors from observed photometry. Grouping galaxies directly by their SED shapes is therefore nearly model-independent when compared to grouping galaxies by their estimated sSFR; thus, this technique produces cleaner groups.

Following \citet{kriek11}, we construct 22 synthetic rest-frame filters equally spaced in $\log\lambda$ between 1,250\AA\ and 40,000\AA. We use EAZY \citep{brammer08} to measure the rest-frame flux of each galaxy in each rest-frame filter that falls within the observed wavelength coverage of the galaxy. For each pair of galaxies in our sample, we calculate a similarity score:
\begin{equation}
    b_{12} = \frac{\Sigma(f_1 - a_{12} f_2)^2}{\Sigma(f_2)^2},
\end{equation}
where $f_1$ is $f_\lambda$ for galaxy 1, $f_2$ is $f_\lambda$ for galaxy 2, $a_{12}$ is a scaling factor given by 
\begin{equation}
    a_{12} = \frac{\Sigma f_1 f_2}{\Sigma f_2^2},
\end{equation}
and the sums are performed over all rest-frame filters that the two galaxies have in common. Two galaxies are considered analogs if their similarity score $b_{12} \le 0.07$. 

We find the galaxy in our sample with the largest number of high-S/N analogs. We quantify this as the galaxy where $N_{\rm{analogs}} \times (\rm{S/N}_{\rm{analogs}})^2$ is maximized. We include this S/N factor to prevent forming very large groups composed of noisy, low-S/N star-forming galaxies. After finding this first primary galaxy, we remove it and its analogs from our sample. 

We repeat this process until there are no galaxies remaining that have more than 19 analogs. Because some galaxies removed in early groups may be more similar to primary galaxies found later in the algorithm, after identifying the groups we re-assign analog galaxies to the group whose primary they are the most similar to. After this process, a total of 5,840 galaxies (83\% of the parent sample) are classified into 26 groups.
To perform the rest of our analysis, we wish to have a relatively small number of groups, each of which has a large number of galaxies. To achieve this goal, we aggregate any groups of galaxies whose primaries have a similarity score of $b_{12} < 0.1$. After this process, we have 16 groups of galaxies. 

The contours in Figure~\ref{fig:ingroup} show the distribution of the full \citet{suess19a} parent sample in both size-mass space and $U-V$ vs $V-J$ (`$UVJ$') color-color space. Points indicate galaxies that were {\it not} classified into a group. The 17\% of galaxies that are not included in a group are spread fairly uniformly across both size-mass and $UVJ$ space. 
The fraction of galaxies not classified into a group is highest in the highest-redshift bin ($2.0 \le z \le 2.5$), likely because these galaxies tend to be fainter and thus have lower signal-to-noise ratios and less reliable photometry. 

\begin{figure*}[ht]
    \centering
    \includegraphics[width=.9\textwidth]{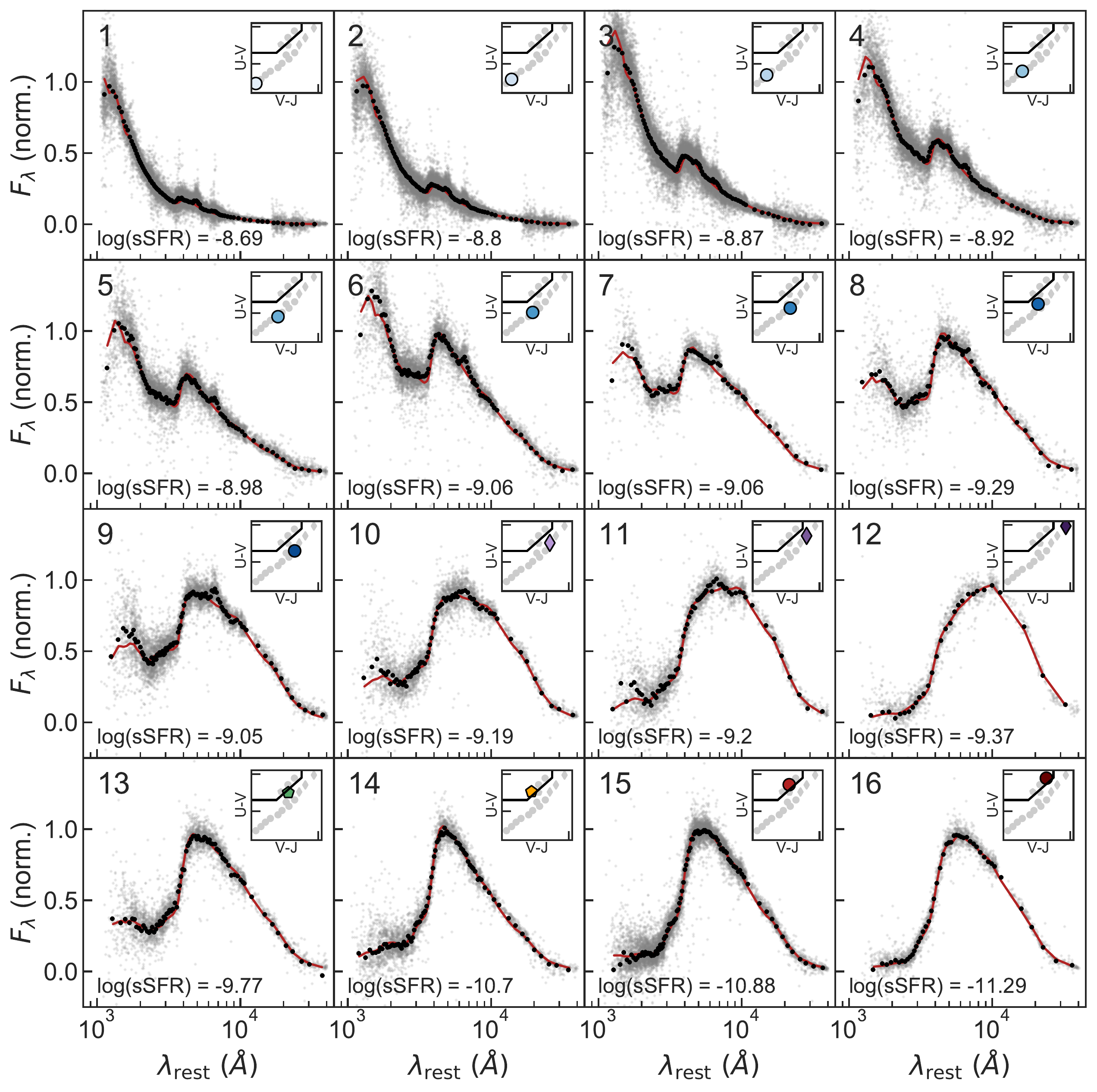}
    \caption{Composite SEDs for each group, ordered by their location in $UVJ$ space. Individual galaxy data points are shown in grey, the binned composite SED is shown in black, and the best-fit stellar population synthesis model (binned to the same resolution as the composite SED) is shown in red. The specific star formation rate of the best-fit model is listed in the lower left of each panel; other best-fit values are shown in Table~\ref{table:compSEDs}. The inset at the top right of each panel shows the location of the composite SED in $UVJ$ space (see Figure~\ref{fig:fullSample} for a larger $UVJ$ diagram with axis labels). The colors and symbols used to represent each group in the $UVJ$ inset remain the same throughout the remainder of this paper.}
    \label{fig:compSEDs}
\end{figure*}

\begin{table}[]
\caption{Best-fit stellar population synthesis fitting results for the composite SED of each group of galaxies.}
\label{table:compSEDs}
\begin{tabular}{llllll}
\hline
Group & $N_{\rm{galaxies}}$ & $\log{\tau}$ & log(age) & \av & log(sSFR) \\ \hline
1            & 1060                & 8.3                   & 8.7                       & 0.15  & -8.69     \\
2            & 1363                & 8.3                   & 8.75                      & 0.3   & -8.8      \\
3            & 1170                & 8.5                   & 8.9                       & 0.45  & -8.87     \\
4            & 535                 & 8.3                   & 8.8                       & 0.6   & -8.92     \\
5            & 295                 & 8.5                   & 8.95                      & 0.75  & -8.98     \\
6            & 233                 & 8.7                   & 9.1                       & 0.85  & -9.06     \\
7            & 88                  & 9.1                   & 9.3                       & 1.05  & -9.06     \\
8            & 100                 & 8.4                   & 9.0                       & 1.0   & -9.29     \\
9            & 168                 & 8.9                   & 9.2                       & 1.35  & -9.05     \\
10            & 104                 & 8.6                   & 9.1                       & 1.55  & -9.19     \\
11           & 97                  & 8.3                   & 8.9                       & 1.95  & -9.2      \\
12           & 42                  & 8.1                   & 8.8                       & 2.35  & -9.37     \\
13           & 110                 & 8.3                   & 9.05                      & 0.95  & -9.77     \\
14           & 123                 & 8.0                   & 8.95                      & 0.55  & -10.7     \\
15           & 285                 & 8.2                   & 9.15                      & 0.5   & -10.88    \\
16           & 67                  & 8.2                   & 9.2                       & 0.7   & -11.29    \\ \hline
\end{tabular}
\tablecomments{Uncertainties in these derived properties are dominated not by formal uncertainties (calculated to be $<0.05$~dex from the stellar population fits), but by systematic uncertainties. These systematic uncertainties are expected to be $\sim0.2$~dex \citep[e.g.,][]{marchesini09,muzzin13}.}
\end{table}

\begin{figure}[ht]
    \centering
    \includegraphics[width=.49\textwidth]{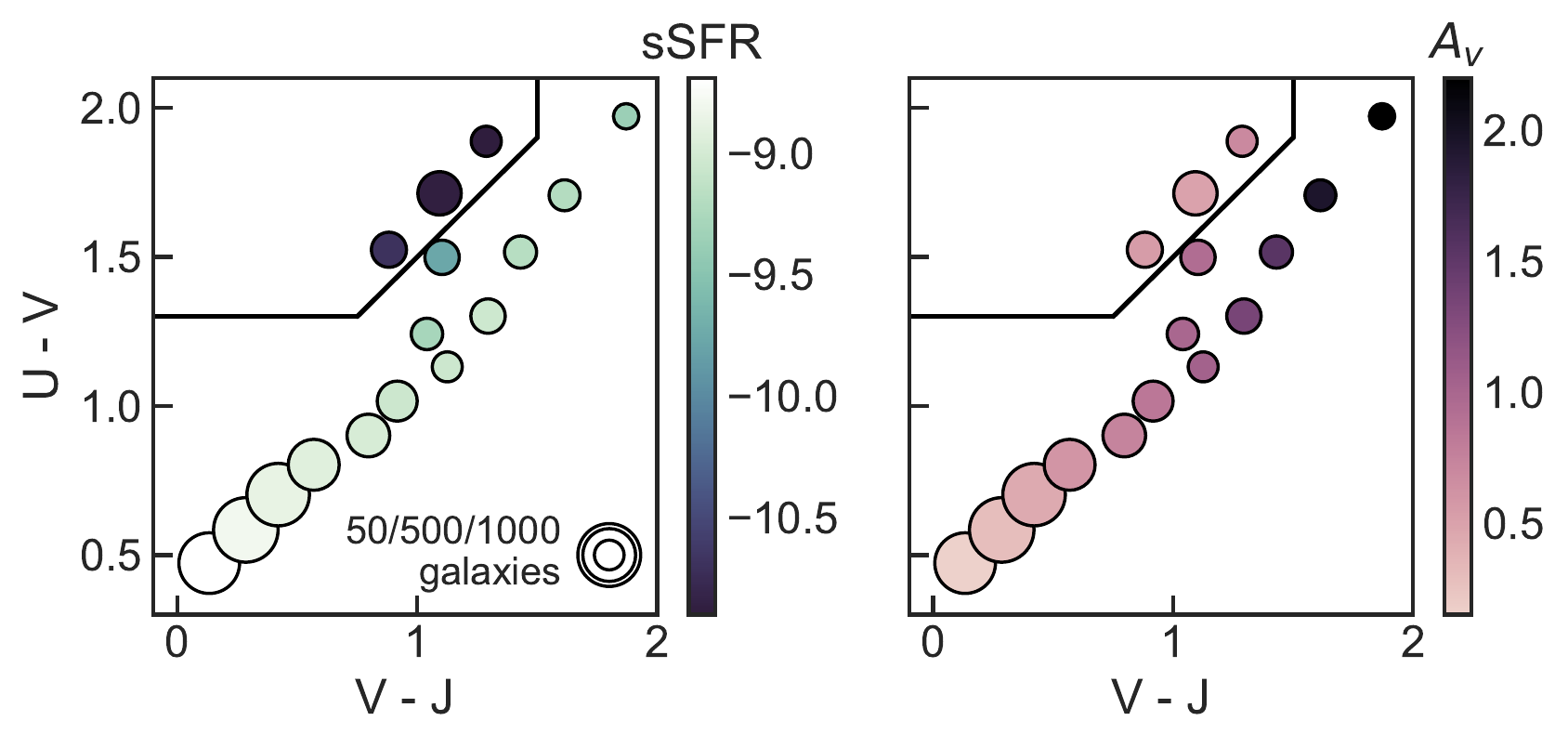}
    \caption{$UVJ$ location of all groups, colored by sSFR (left) or A$_v$ (right). The area of the symbol is proportional to the number of galaxies in the group. sSFR decreases slightly moving up the $UVJ$ star-forming sequence, and decreases dramatically moving from the star-forming to quiescent region of $UVJ$ space. A$_v$, however, shows most variation moving up the $UVJ$ star-forming sequence, with galaxies in the upper-right corner being significantly dustier than those in the lower-left corner.}
    \label{fig:uvj_sed}
\end{figure}

\begin{figure*}
    \centering
    \includegraphics[width=\textwidth]{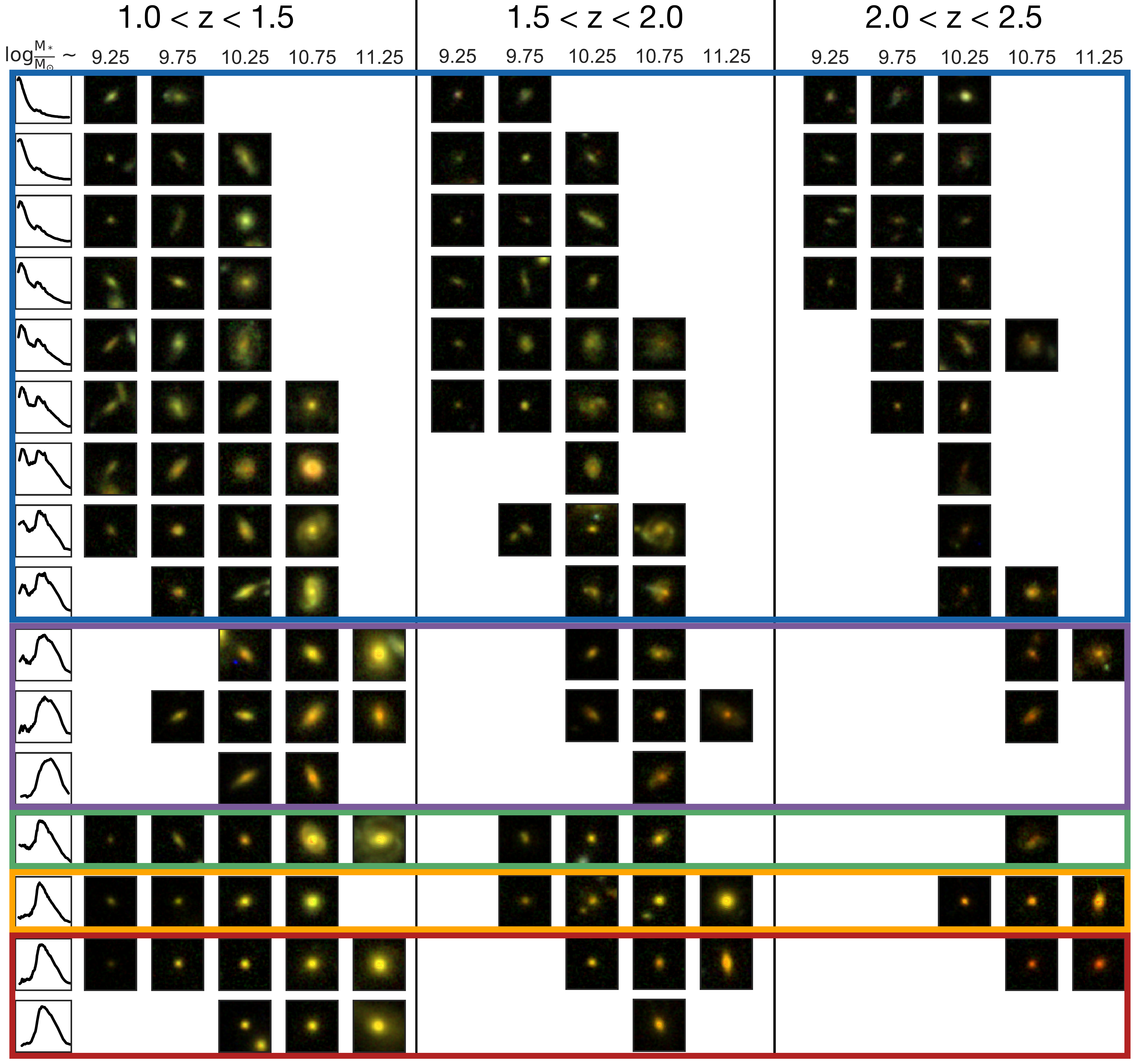}
    \caption{Example color images of galaxies in each group across mass and redshift (from F160W, F125W, and F814W CANDELS imaging). Cutouts are $3"$ per side; this corresponds to $\sim25$~kpc across our full redshift range. No image is shown if a group does not contain any galaxies in a given mass and redshift bin. Unobscured star-forming galaxies are highlighted in blue, dusty star-forming galaxies in purple, green valley galaxies in green, post-starburst galaxies in yellow, and quiescent galaxies in red. Galaxy structure clearly depends on both SED shape, redshift, and mass, as explored in detail in the remainder of this paper.}
    \label{fig:images}
\end{figure*}

Figure \ref{fig:compSEDs} shows the composite SED of each group (black points) as well as the observed data for each individual galaxy in the group (small grey points). We create these composite SEDs following the \citet{kriek11} technique: the observed SEDs of all galaxies in each group are de-redshifted, scaled to the same arbitrary flux normalization using the $a_{12}$ scaling factor, then median binned. Scaled data points are binned such that there are at least 50 individual points in each composite SED point and no more than 100 total points in the composite SED. Error bars on the composite SED points are computed by bootstrap resampling the individual galaxy SEDs.  
We also construct an effective filter response curve for each composite SED point by de-redshifting the true filter response curves for each data point in the median point, then normalizing and adding them. 

We use these effective filter response curves to fit each composite SED with the FAST stellar population synthesis fitting code \citep{kriek09}. 
In these fits, we fix the redshift of the composite SED to $z=0$ and assume the \citet{bc03} stellar population library, a \citet{chabrier03} initial mass function, the \citet{kriek13} dust attenuation law, and a delayed exponential star formation history. We mask points in the composite SED within 400$\mathrm{\AA}$ of the H$\alpha$, H$\beta$, and [OIII]$\lambda$5007 lines, as well as points within 75$\mathrm{\AA}$ of the [OII]$\lambda$3727 line; these points are often visibly contaminated by line emission, which is not included in the FAST templates. The red lines in Figure~\ref{fig:compSEDs} show the best-fit model to each composite SED, binned to the same resolution as the composite SED. The best-fit age, star-formation timescale, dust attenuation, and sSFR for each composite SED is listed in Table~\ref{table:compSEDs}; the sSFR is also shown in the bottom left of each panel in Figure \ref{fig:compSEDs}. {Figure~\ref{fig:uvj_sed} shows the sSFR and A$_v$ of all groups in $UVJ$ space.} These quantities are mass-independent, and therefore are not affected by the arbitrary normalization of the composite SEDs. We note that the masses of galaxies in each group do systematically differ, and generally increase as we move through Figure~\ref{fig:compSEDs} and sSFR decreases; these differences are explored in detail in later sections of this paper. 
The inset in the upper right of each panel in Figure~\ref{fig:compSEDs} shows the location of each group's composite SED in $UVJ$ space; the line in this inset demarcates the quiescent region \citep[again, using the definition from][]{whitaker12}. 

{Figures~\ref{fig:compSEDs} \& \ref{fig:uvj_sed}} show the incredible diversity of rest-frame SED shapes present in our sample. Because groups 1-13 all lie in the star-forming section of the $UVJ$ diagram, they are traditionally all grouped together to study the structural relations for star-forming galaxies. However, {Figures~\ref{fig:compSEDs} \& \ref{fig:uvj_sed}} show that these galaxies have vastly different properties: the star-formation timescale, age, dust content, {\it and} sSFR vary dramatically from group to group. Figure~\ref{fig:compSEDs} also allows us to see a more gradual shift from star-forming galaxies to quiescent ones: there are a number of `intermediate' SED shapes that have shapes between highly star-forming and fully quenched. 

To guide the eye, we plot each galaxy's SED type using a different color and marker for the remainder of this paper. We plot star-forming galaxies as blue points; the shade describes how far up the $UVJ$ star-forming sequence the group lies. We identify three groups (10, 11, and 12) of very dusty star-forming galaxies which have A$_v>1.5$. These galaxies are shown using purple diamonds. Fully quiescent galaxies, with sSFR~$<-10.8$, are shown as red points. 
Finally, groups 13 and 14 are located very close to the boundary of the star-forming and quiescent sections of the $UVJ$ diagram. These `transitioning' galaxies are shown with pentagons. Group 13 consists of galaxies just on the star-forming side of the $UVJ$ diagram that have some UV flux, relatively high dust obscuration, and fairly low specific star formation rates. We refer to these galaxies as `green valley' galaxies and plot them with a green pentagon \citep[see, e.g.,][]{patel11,fang18,gu18}. Group 14 consists of galaxies at the lower-left corner of the $UVJ$ quiescent region with low UV fluxes, low specific star formation rates, short star-formation timescales, and a sharp Balmer break. These galaxies are often referred to as `post-starburst' \citep[e.g.,][]{whitaker12_psb,belli19}; we plot them with a yellow pentagon.

We calculate the mass completeness of each group using a technique similar to \citet{quadri12} and \citet{tomczak14}: we scale the $K_s$-band magnitude and mass of each galaxy in the group that has a signal-to-noise value close to our cutoff ($20 \le \rm{ S/N } \le 50$) down to the flux completeness of the survey \citep{straatman16}. We take the 90th percentile of these masses as the mass completeness of the group. We calculate this mass completeness in three redshift bins: $1.0 \le z < 1.5$, $1.5 \le z < 2.0$, and $2.0 \le z \le 2.5$. The results that follow will also be broken into these same redshift intervals.

Figure~\ref{fig:images} shows example images of galaxies in each group as a function of mass and redshift. These postage stamps demonstrate that galaxy structure and morphology clearly depends on both SED shape, mass, and redshift.

\section{Results}
\label{sec:results}

\begin{figure*}[ht]
    \centering
    \includegraphics[width=.8\textwidth]{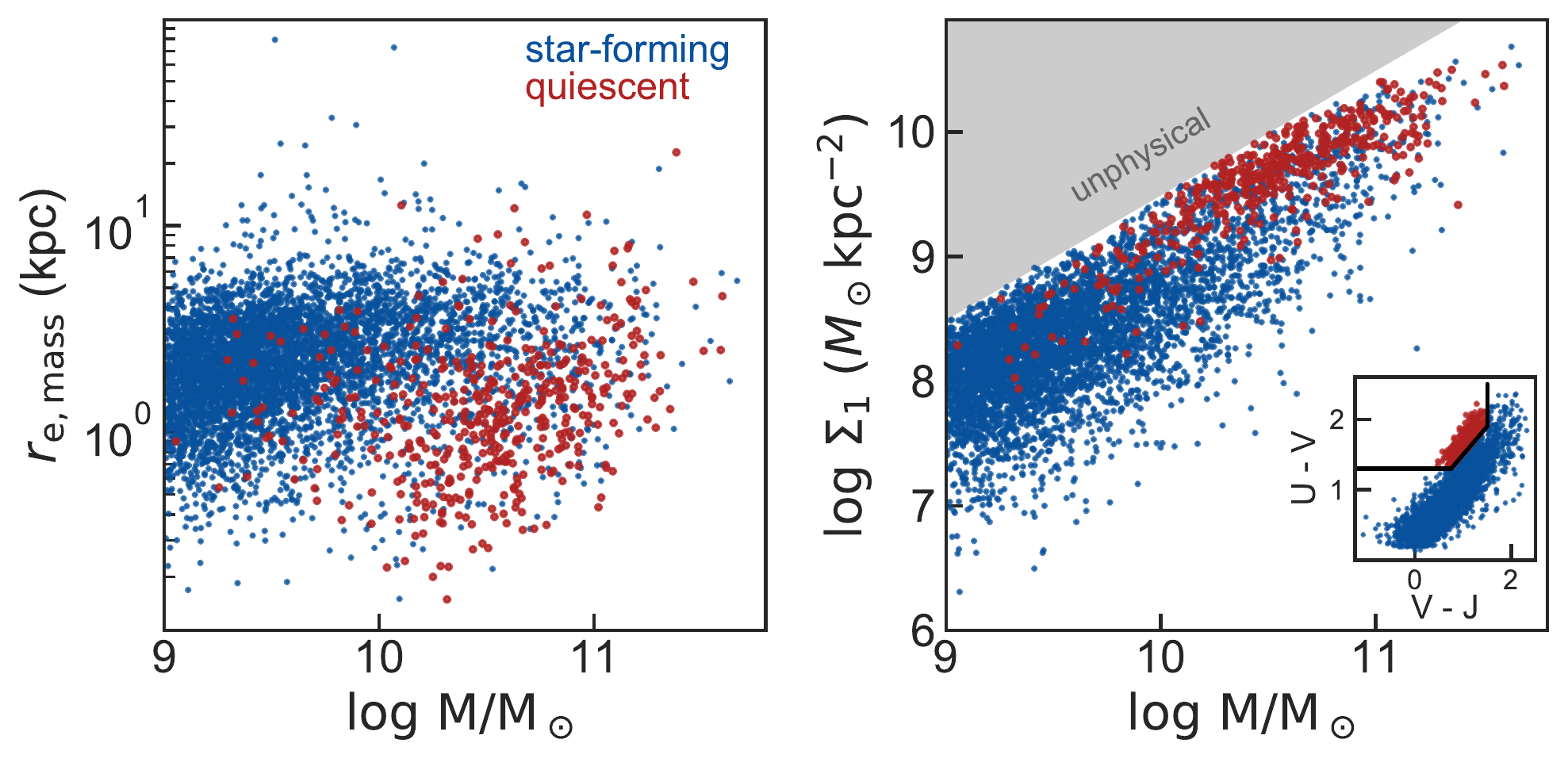}
    \caption{Our full sample in size-mass (left) and \sig-mass (right) space. The inset shows the full sample in $UVJ$ space. Star-forming galaxies are shown in blue, and quiescent galaxies in red. Because the median half-mass radii of galaxies do not evolve rapidly over the $1.0\le z\le 2.5$ interval we study here \citep{suess19a,suess19b}, for brevity we show all redshifts in the same figure. The grey shaded region in \sig-mass space shows the unphysical regime, where $>100\%$ of the galaxy's mass lies within the central kiloparsec. While star-forming and quiescent galaxies follow relatively tight relations with similar slopes in \sig-mass space, they have more scatter and clearly different slopes in size-mass space. The slope of the star-forming size-mass relation is nearly flat, consistent with the minimal size evolution we observe over this redshift regime.}
    \label{fig:fullSample}
\end{figure*}

Over the past decade, many studies have explored the distribution of galaxies both in the size-mass plane \citep[e.g.,][]{shen03,vanderwel14,omand14,whitaker17,mowla18,suess19a} and the \sig-mass plane \citep[e.g.,][]{fang13,barro17,whitaker17,lee18,woo19}. In this paper, we pursue two new avenues to advance this legacy of structural studies. First, our measurements are based on mass profiles; this removes potential biases due to radial color gradients \citep[e.g.,][]{szomoru13,chan16,suess19a,suess19b,suess20}. Second, instead of studying the structures of $UVJ$-selected star-forming and quiescent galaxies, we study the structures of a large number of galaxy groups identified by their similar SED shapes (Section~\ref{sec:sed_groups}). 

Before dissecting these structural relations, we consider how the traditional star-forming and quiescent size-mass and \sig-mass relations change when we use color gradient-corrected mass profiles. 
Figure~\ref{fig:fullSample} shows the the structural properties of our sample, with $UVJ$-selected star-forming galaxies in blue and $UVJ$-selected quiescent galaxies in red; the inset in the lower right shows the $UVJ$ plane. 
The grey shaded region in the \sig-mass plane shows the unphysical regime where $M_{\rm{r<1kpc}} > M_{\rm{tot}}$. For brevity, we show the full $1.0\le z\le2.5$ redshift range in one panel; as shown in \citet{suess19a}, galaxy half-mass radii of both star-forming and quiescent galaxies do not evolve significantly in this redshift regime. In agreement with the literature, both star-forming and quiescent galaxies lie on well-defined size-mass relations; quiescent galaxies are smaller at fixed mass, and have a steeper slope in size-mass space than star-forming galaxies \citep[e.g.,][]{shen03,vanderwel14,mowla18}. We also see that the quiescent size-mass relation clearly flattens below $\sim10^{10}M_\odot$ \citep[e.g.,][]{cappellari13,vanderwel14,whitaker17}; we discuss this flattening further in Section~\ref{sec:discuss_qui}. 

Because we use half-mass radii, the overall structural relations we find for star-forming and quiescent galaxies differ from previous studies based on half-light radii. More massive galaxies have stronger color gradients, so the slope of the star-forming size-mass relation is flatter than found by previous studies using half-light radii \citep[e.g.,][]{vanderwel14,mowla18,suess19a}. For star-forming galaxies, this shallower slope can be explained by $A_{\rm{v}}$ gradients: massive star-forming galaxies are dustier \citep[e.g.][]{whitaker17_dust}, and dust profiles in star-forming galaxies at these redshifts tend to be centrally-peaked \citep[e.g.,][]{nelson16_dust,tacchella18}. These $A_{\rm{v}}$ gradients cause more massive star-forming galaxies to have stronger negative color gradients, and thus smaller half-mass radii.
As discussed in \citet{suess19b}, this flatter slope is consistent with the minimal size growth that we observe at these redshifts. In turn, this slow growth agrees with IFU studies that show roughly flat dust-corrected sSFR profiles for star-forming galaxies in this redshift regime \citep[e.g.,][]{tacchella15_science,tacchella15_apj,nelson16}. 

Star-forming and quiescent galaxies also lie on well-defined relations in \sig-mass space. Unlike the size-mass relations, the \sig-mass relations of the two galaxy types have very similar slopes, and the scatter around the relations is smaller. Quiescent galaxies are offset to higher \sig values at fixed mass, effectively tracing out the upper half of the star-forming \sig-mass relation. In agreement with previous studies, the quiescent \sig-mass relation is quite tight, with much less scatter than the quiescent size-mass relation \citep[e.g.,][]{fang13,barro17,chen20}. Again, we find that the star-forming sequence is offset slightly from previous measurements \citep[e.g.,][]{barro17}. This difference can also be ascribed to stronger color gradients in more massive star-forming galaxies, which increase our \sig values for massive star-forming galaxies. 

With this picture of the star-forming and quiescent size-mass and \sig-mass relations in place, we now begin to dissect these structural relations by examining the color gradients, sizes, and \sig values for each group of galaxies shown in Figure~\ref{fig:compSEDs}.

\label{sec:groupResults}

\begin{figure*}[ht]
    \centering
    \includegraphics[width=\textwidth]{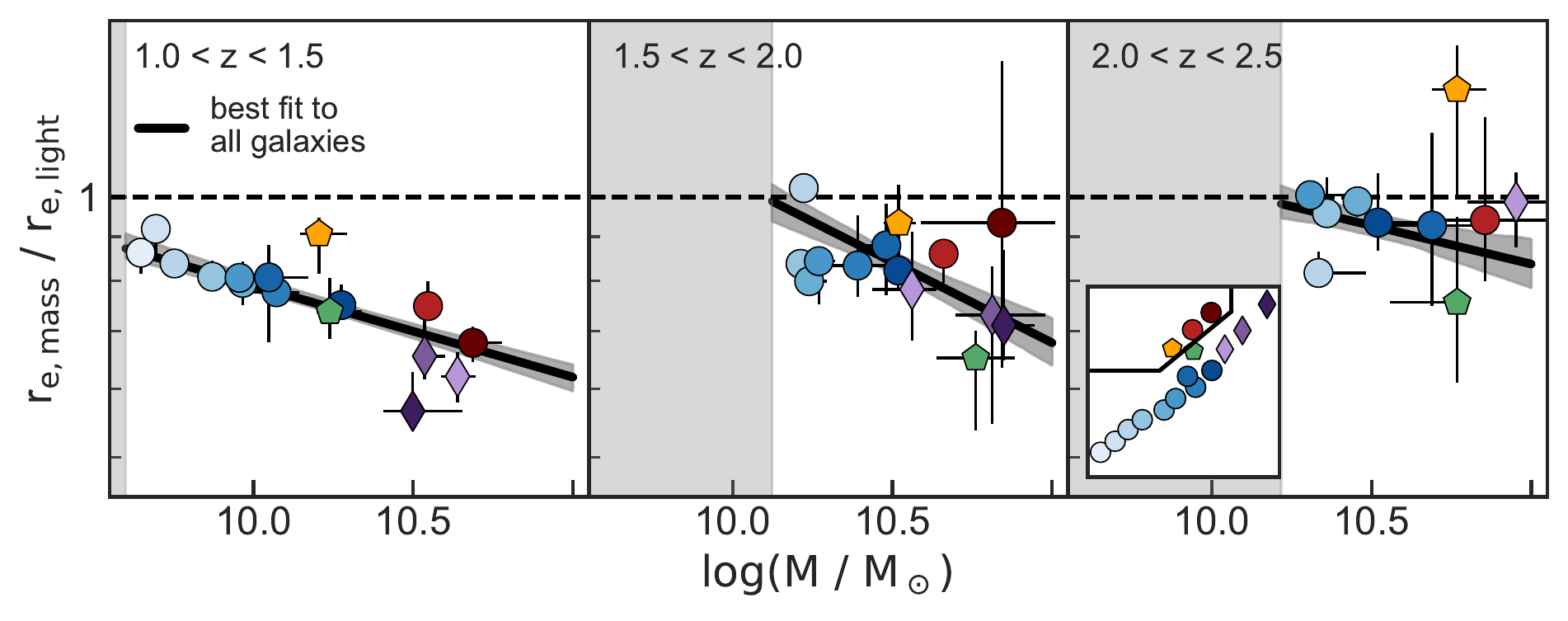}
    \caption{Median color gradient strength as a function of median stellar mass for each galaxy group. Each point represents one SED group; the color and symbol used for each group corresponds to its location in $UVJ$ space, as shown in the inset. Error bars on individual points represent the 1$\sigma$ spread in bootstrap resamples of the median $r_{\rm{e, mass}} / r_{\rm{e, light}}$. The black line and grey shaded region show the \citet{suess19a} fit to all galaxies and its $1\sigma$ error. The grey shaded region on the left shows the mass completeness in each redshift interval. Groups have systematically different color gradient strengths, but typically have the expected value for their mass; notable exceptions are the post-starburst galaxies (yellow pentagon), which have weak to flat color gradients, and dusty star-forming galaxies (purple diamonds), which have strongly negative color gradients at low redshift.}
    \label{fig:groupRatio}
\end{figure*}
\subsection{Color gradient strength varies with SED shape}

Color gradient strength depends on stellar mass \citep[e.g.,][]{tortora10,suess19a}, redshift \citep{suess19a,suess19b} and age along the quiescent sequence \citep{suess20}; here, we test whether these systematic color gradient differences extend to the sixteen star-forming, transitional, and quiescent SED types shown in Figure~\ref{fig:compSEDs}. As in \citet{suess19a,suess19b,suess20}, we use the ratio of the galaxy's half-mass and half-light radius to probe color gradient strength. High values of $r_{\rm{e, mass}} / r_{\rm{e, light}}$ indicate positive color gradients, where the center of the galaxy is bluer than the outskirts; low values of $r_{\rm{e, mass}} / r_{\rm{e, light}}$ indicate negative color gradients, where the center of the galaxy is redder than the outskirts; $r_{\rm{e, mass}} / r_{\rm{e, light}} = 1$ indicates no radial color gradient. 

Each point in Figure \ref{fig:groupRatio} shows the median $r_{\rm{e, mass}} / r_{\rm{e, light}}$ value of all galaxies in a group as a function of the group's median stellar mass. Median points exclude galaxies below the mass completeness limit at each redshift (shaded vertical grey region). The symbols and colors of the points correspond to the group's location in $UVJ$ space, as shown in the inset. The solid black line (and  grey region) indicate a best-fit line (and 16-84\% confidence interval) to the trend in color gradient strength as a function of stellar mass for all galaxies in the sample \citep{suess19a}. 

Figure \ref{fig:groupRatio} shows that different groups of galaxies have systematically different color gradients. In general, groups with higher masses and lower specific star formation rates tend to have more strongly negative color gradients. 
These variations in color gradient strength with SED shape are generally consistent with the \citet{suess19a} relation between color gradient strength and stellar mass for all galaxies. 
The most notable outliers from this trend are the post-starburst galaxies, shown in yellow. These young quiescent galaxies have systematically {\it weaker} color gradients than expected given their stellar mass. As discussed in \citet{suess20} and Section~\ref{sec:discuss_qui}, these flat color gradients are consistent with post-starburst galaxies being the result of a ``fast" quenching process that requires structural change. The other possible outliers are dusty star-forming galaxies, which at $1.0\le z\le2.5$ appear to have stronger color gradients than expected from their stellar mass. We discuss the color gradients of dusty star-forming galaxies in detail in Section~\ref{sec:discuss_qui}.

These variations in color gradient strength with SED shape systematically alter the half-light radii of different types of galaxies. It is therefore {\it essential} that we account for these varying color gradients when dissecting the size-mass and \sig-mass relations. In this study, we account for these varying color gradients by examining the half-{\it mass} radii of galaxies. 

\subsection{Galaxies lie in distinct parts of size-mass \& \sig-mass space}
\begin{figure*}
    \centering
    \includegraphics[width=\textwidth]{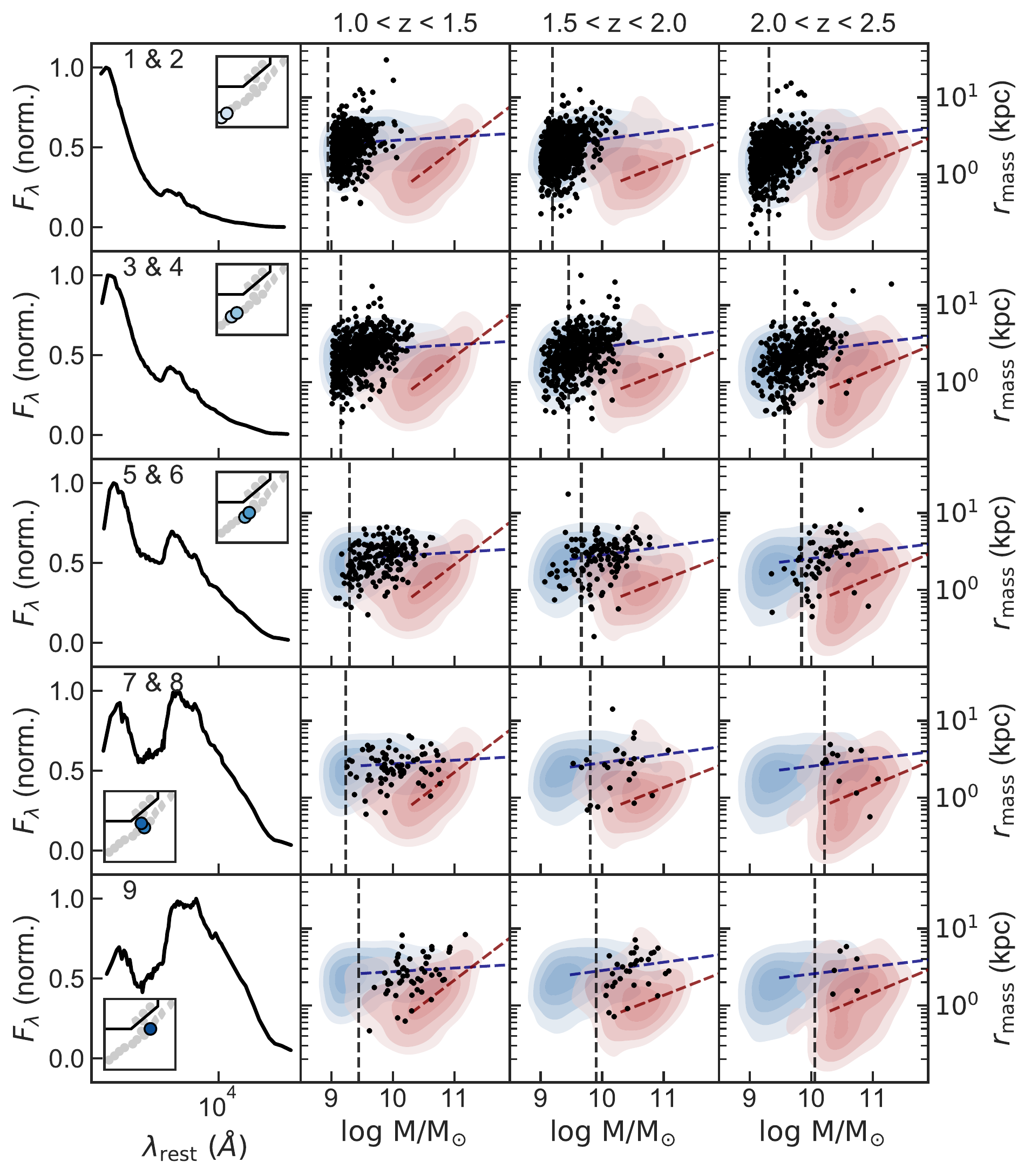}
    \caption{Half-mass radius versus mass for each group of galaxies, ordered by decreasing sSFR. Left: composite SED of the group, normalized to a peak flux of 1.0.  
    The inset shows the group's location in $UVJ$ space. The next three panels show $r_{\rm{e,mass}}$ as a function of stellar mass for three different redshift ranges. Black points show the galaxies in that group; for comparison, the total sample of star-forming (quiescent) galaxies is shown as blue (red) contours. The contours enclose 25, 50, 75 and 98\% of the full sample at each redshift. The vertical grey dashed line shows the mass completeness for each group. The blue and red dashed lines show the $r_{\rm{e,mass}}-M_*$ relations from \citet{suess19a}.}
\end{figure*}
\begin{figure*}
    \ContinuedFloat
    \centering
    \includegraphics[width=\textwidth]{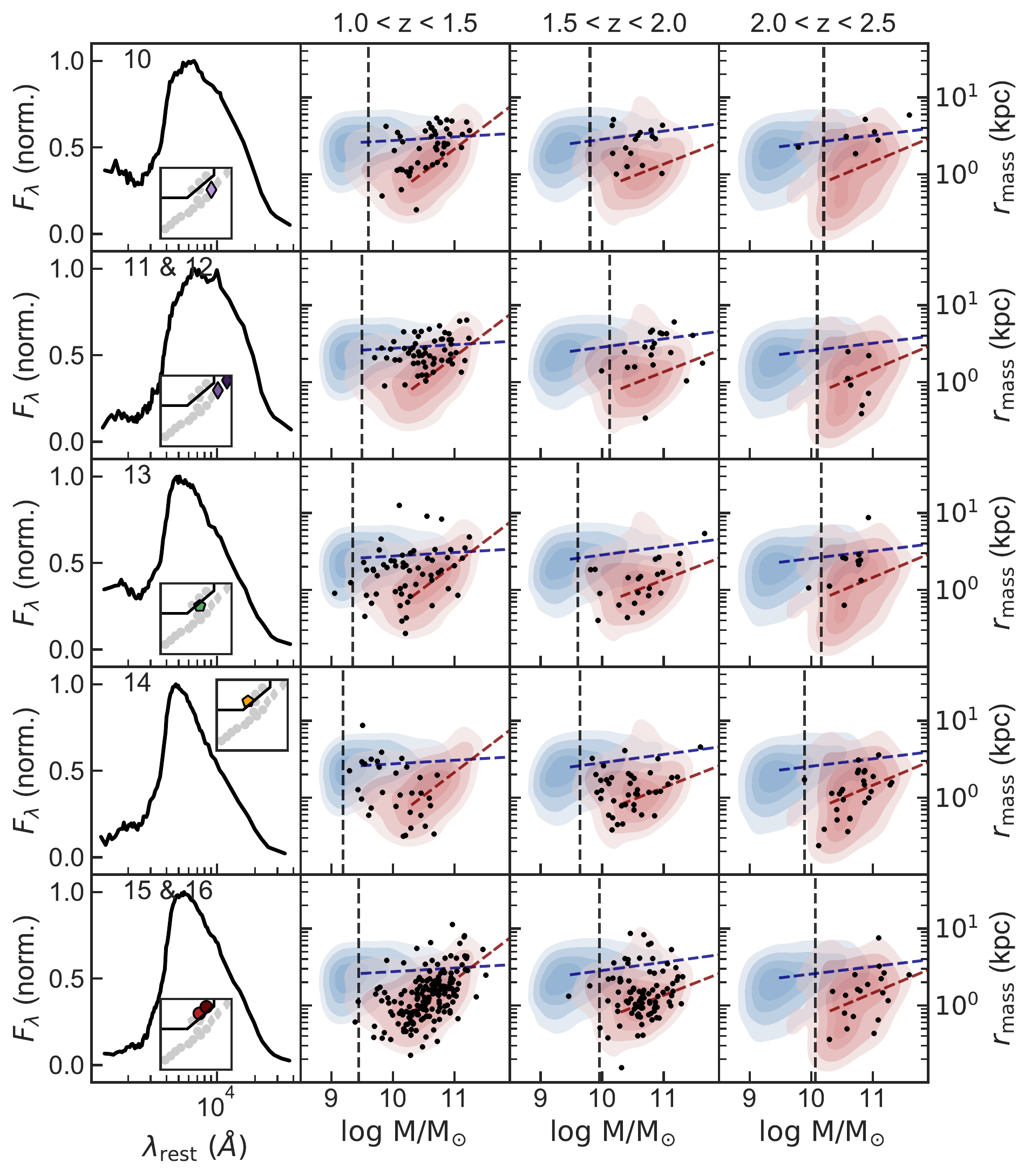}
    \caption{Continuation of previous figure.}
    \label{fig:groupMassSize}
\end{figure*}

\begin{figure*}
    \centering
    \includegraphics[width=\textwidth]{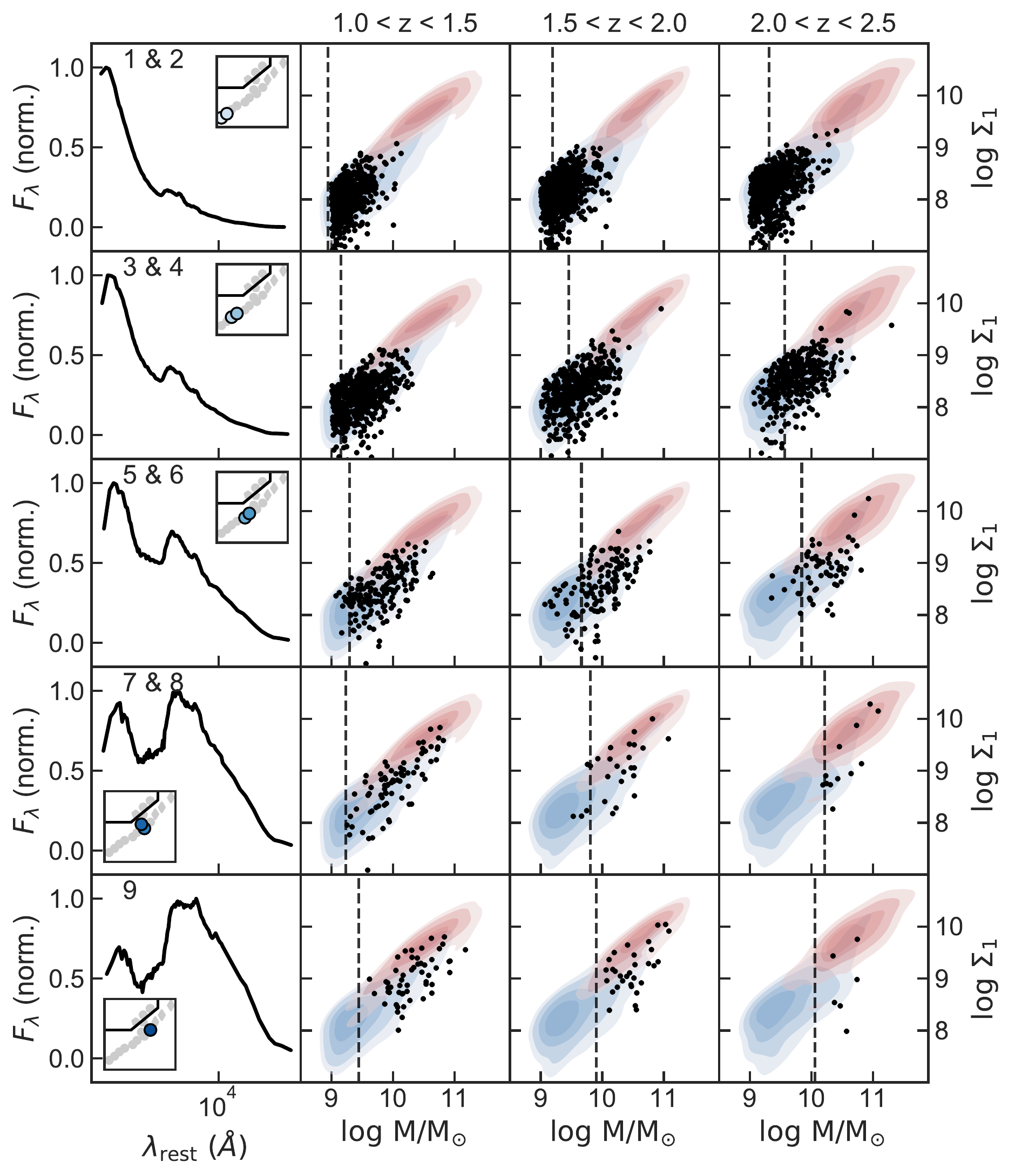}
    \caption{Same as Figure~\ref{fig:groupMassSize}, but showing \sig versus mass for each group.}
\end{figure*}
\begin{figure*}
    \ContinuedFloat
    \centering
    \includegraphics[width=\textwidth]{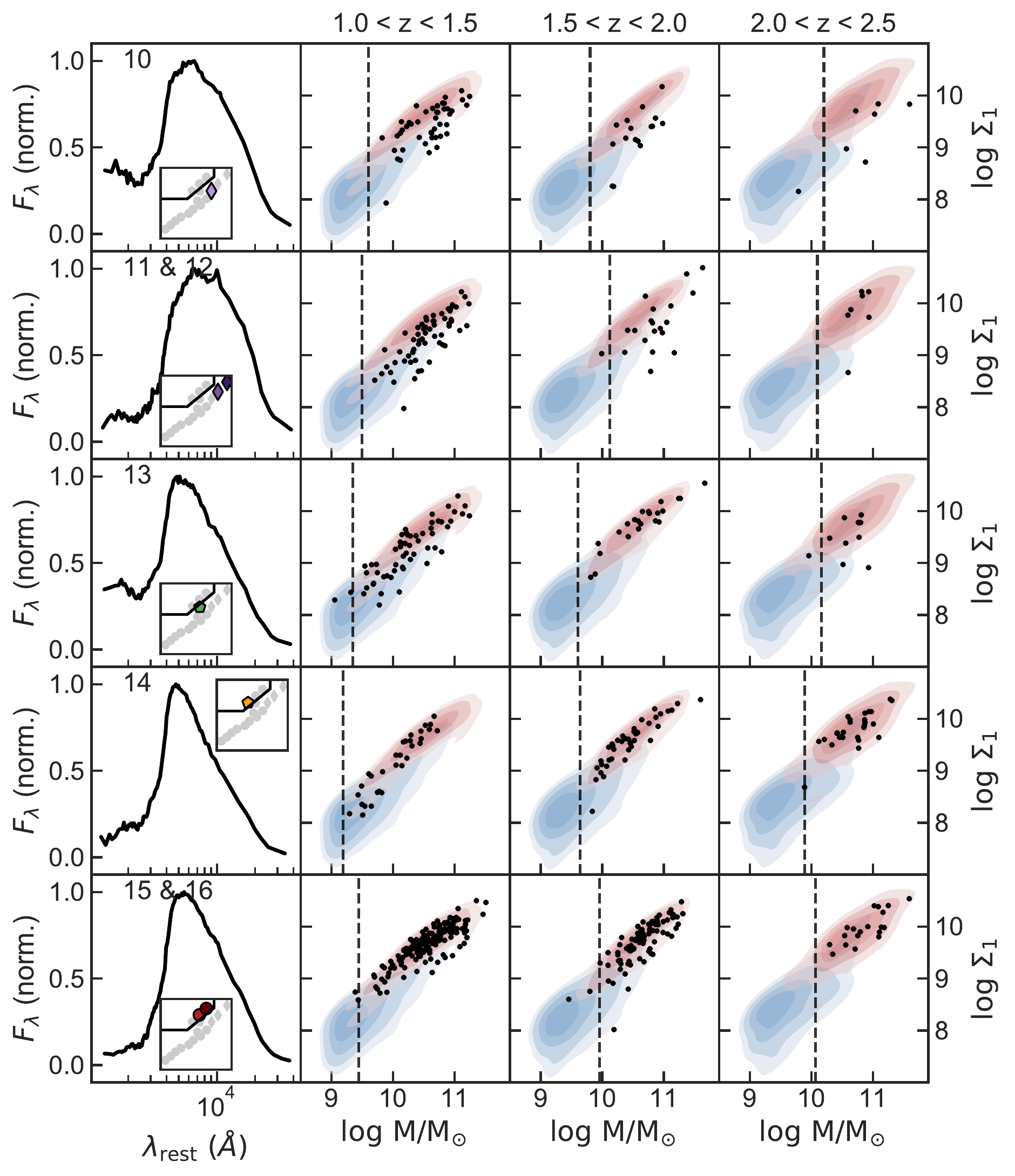}
    \caption{Continuation of previous figure.}
    \label{fig:groupMassSig}
\end{figure*}

We now turn towards understanding where the galaxies in each group lie in size-mass and \sig-mass space, corrected for color gradients. Figure \ref{fig:groupMassSize} shows half-mass radius as a function of stellar mass for the galaxies in each group. The leftmost column shows the composite SED of the group, as well as the group's location in $UVJ$ space. The right columns show three different redshift slices of the size-mass diagram. The black points indicate the masses and half-mass radii of the galaxies with that particular SED type. The blue and red contours show the full galaxy population, divided into star-forming and quiescent groups using the $UVJ$ classification of \citet{whitaker12}. The blue and red dashed lines represent the $r_{\rm{mass}} - M_*$ relations from \citet{suess19a}, calculated from the same galaxy sample and size measurements used in this paper. 
The size-mass contours and best-fit relations are the same in each column, and are shown to indicate how each group compares to the traditional `blue vs. red' view of the galaxy size-mass diagram.  
For brevity, several groups which have very similar composite SEDs and locations in size-mass space are shown in the same row of Figure \ref{fig:groupMassSize}. For these combined groups, we show only one composite SED (which includes all galaxies from both groups), but show both $UVJ$ points.

In general, stellar mass increases and sSFR decreases as we move down the rows of Figure \ref{fig:groupMassSize} and march up the $UVJ$ diagram. The mass completeness limits also increase with decreasing sSFR, but star-forming galaxies are less massive than quiescent galaxies even when we consider a very conservative mass cut. This increase in mass with decreasing sSFR is expected: it is a natural consequence of the shallow slope of the star-forming main sequence \citep[e.g.,][]{brinchmann04,noeske07,daddi07,gonzalez10,whitaker12}.

Figure \ref{fig:groupMassSize} shows that each group of galaxies occupies a distinct and fairly localized region of size-mass space. Breaking the size-mass plane into just two relations, star-forming and quiescent, is an oversimplification of the complexity we see in Figure~\ref{fig:groupMassSize}: both the star-forming and quiescent regions of size-mass space are comprised of multiple different galaxy groups with distinct SED shapes and different sSFRs. 
The star-forming region in particular is populated by many different groups, discussed in detail in Section~\ref{sec:discuss_sf}.

Additionally, we find that there are transitional phases where the galaxies lie between the star-forming and quiescent size-mass relations. These transitional phases include post-starburst galaxies (yellow pentagon), green valley galaxies (green pentagon), and high-mass dusty star-forming galaxies (purple diamonds). Interestingly, the sizes of these transitional galaxies clearly depend on redshift. In Section \ref{sec:discuss_qui} we investigate these groups further and discuss how they may represent different pathways to galaxy quenching. 

Figure~\ref{fig:groupMassSig} replicates Figure~\ref{fig:groupMassSize}, but showing where each group lies in \sig-mass space instead of size-mass space. Contours again show the full sample divided into quiescent (red) and star-forming (blue) using a $UVJ$ cut. Individual groups of star-forming galaxies trace out the overall \sig-mass relation up to higher and higher masses as sSFR decreases. Dusty star-forming galaxies appear to be slightly offset to lower \sig values, tracing out the high-mass regime below the bulk of the quiescent population; we discuss the implications of these results in detail in Section~\ref{sec:discuss_qui}. Post-starburst galaxies have high \sig values consistent with the quiescent population, in agreement with our previous results \citep{suess20}. Green valley galaxies also have relatively high \sig values, especially at $z>1.5$.

Figures \ref{fig:groupMassSize} \& \ref{fig:groupMassSig} demonstrate the need to move beyond classifying galaxies into just two groups, star-forming {\it or} quiescent. By creating such broad categories, we are averaging over a huge amount of interesting behavior. Additionally, it is only by looking at smaller groups of galaxies that we can start to build up an understanding of how galaxies evolve {\it through} these spaces-- an understanding that may translate to a greater knowledge of the physical mechanisms responsible for both the mass assembly histories of galaxies and the quenching process.

The remainder of this paper will be devoted to summarizing, discussing, and contextualizing the data shown in Figures \ref{fig:groupMassSize} \& \ref{fig:groupMassSig}.

\subsection{Median sizes and \sig values as a function of sSFR}
\label{sec:sizeAll}

\begin{figure*}
    \centering
    \includegraphics[width=.75\textwidth]{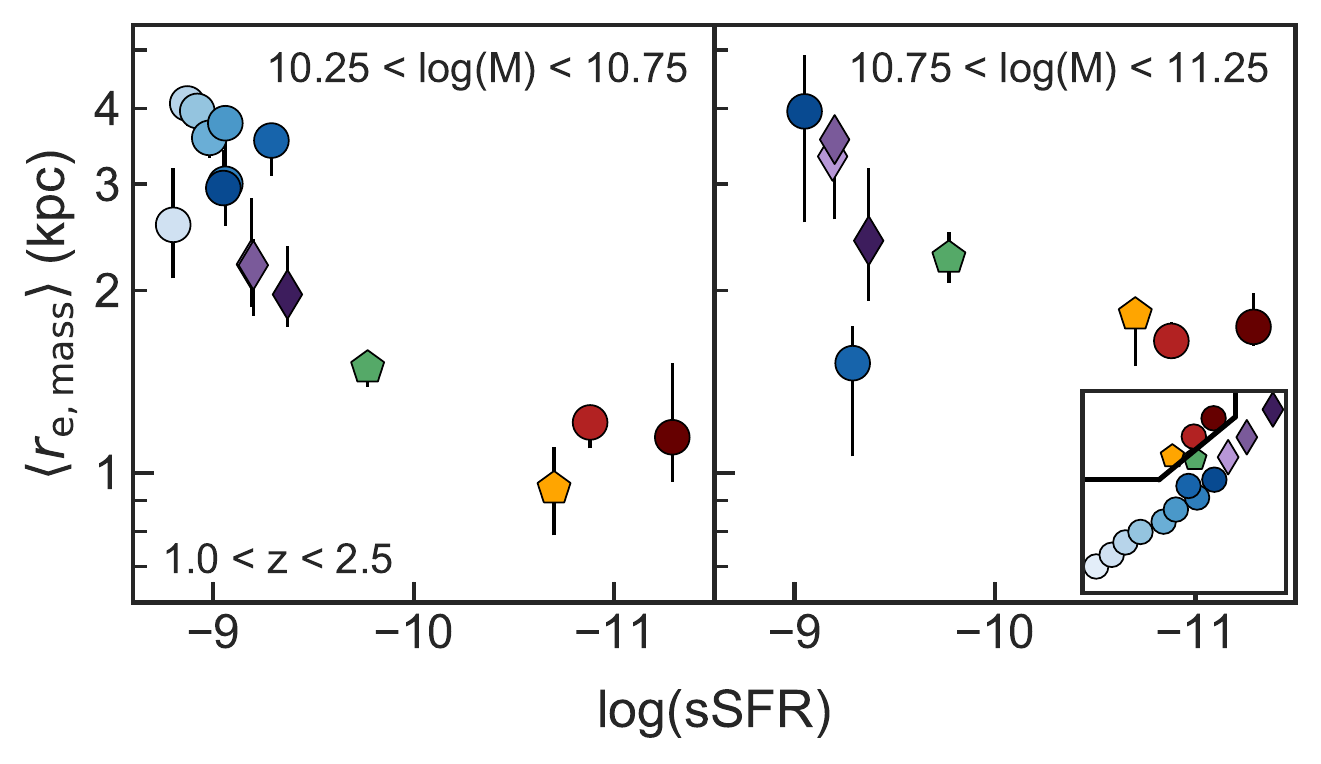}
    \caption{Median half-mass radius as a function of sSFR for each group. Each panel represents a different mass range, $10.25 < \log{\rm{M}} < 10.75$ on the left and $10.75 < \log{\rm{M}} < 11.25$ on the right. These mass ranges are chosen to be above the maximum mass completeness limit for all galaxy types at all redshifts. Error bars on the median mass show the 16-84\% range of 1000 bootstrap resamples of the median half-mass radius. A median point is only shown for each group if there are more than 5 galaxies in a given group and mass range. The median sizes of galaxies decrease gradually as sSFR decreases. Transitioning galaxies, especially green valley galaxies, have intermediate sizes.}
    \label{fig:meanR_ssfr}
\end{figure*}

We begin distilling the size-mass data shown in Figure \ref{fig:groupMassSize} by considering the median size of the galaxies in each group. We compute these median sizes in two mass bins, $10.25 < \log{\rm{M_*/M_\odot}} < 10.75$ and $10.75 < \log{\rm{M_*/M_\odot}} < 11.25$. 
Both mass bins are above the mass completeness limit for all groups at all redshifts. Choosing narrow mass bins also allows us to compute median half-mass radii without taking into account the slope of the size-mass relation. 
Figure \ref{fig:meanR_ssfr} shows median size as a function sSFR for both mass bins. Because the half-mass radii of both star-forming and quiescent galaxies do {not} show significant redshift evolution between $z=1$ and $z=2.5$ \citep{suess19a}, we do not split Figure~\ref{fig:meanR_ssfr} by redshift. We have verified that the interpretation of this figure remains unchanged if we do split the sample into multiple redshift ranges.

In agreement with previous studies, we find that quiescent galaxies are smaller on average than star-forming galaxies in both mass bins \citep[e.g.,][]{shen03, vanderwel14, mowla18}. 
However, there is not a sudden jump in galaxy size at some sSFR: instead, the median size of each galaxy group decreases {\it smoothly} with decreasing sSFR. We find that the sizes of star-forming galaxies decrease by $\sim0.2$~dex as sSFR decreases. This is a slightly stronger trend between size and sSFR than found by \citet{whitaker17}, likely due to the fact that we correct for $A_{\rm{v}}$ gradients \citep[expected to be stronger in higher-mass and lower-sSFR star-forming galaxies due to their increased dust content, e.g.][]{magnelli09,murphy11,bourne17,whitaker17_dust}.
Our results also quantitatively agree with predictions from simulations, where the sizes of star-forming galaxies tend to decrease with decreasing sSFR \citep[e.g.,][]{furlong17,genel18}. Furthermore, we find that transitional galaxy groups--- especially green valley galaxies--- have median sizes between those of star-forming and quiescent galaxies \citep[in agreement with previous studies, e.g.][]{mendez11,yano16,wu18}. As in \citet{suess20}, we find that the median sizes of post-starburst galaxies are not significantly smaller than those of older quiescent galaxies at fixed mass.

\begin{figure*}[ht]
    \centering
    \includegraphics[width=\textwidth]{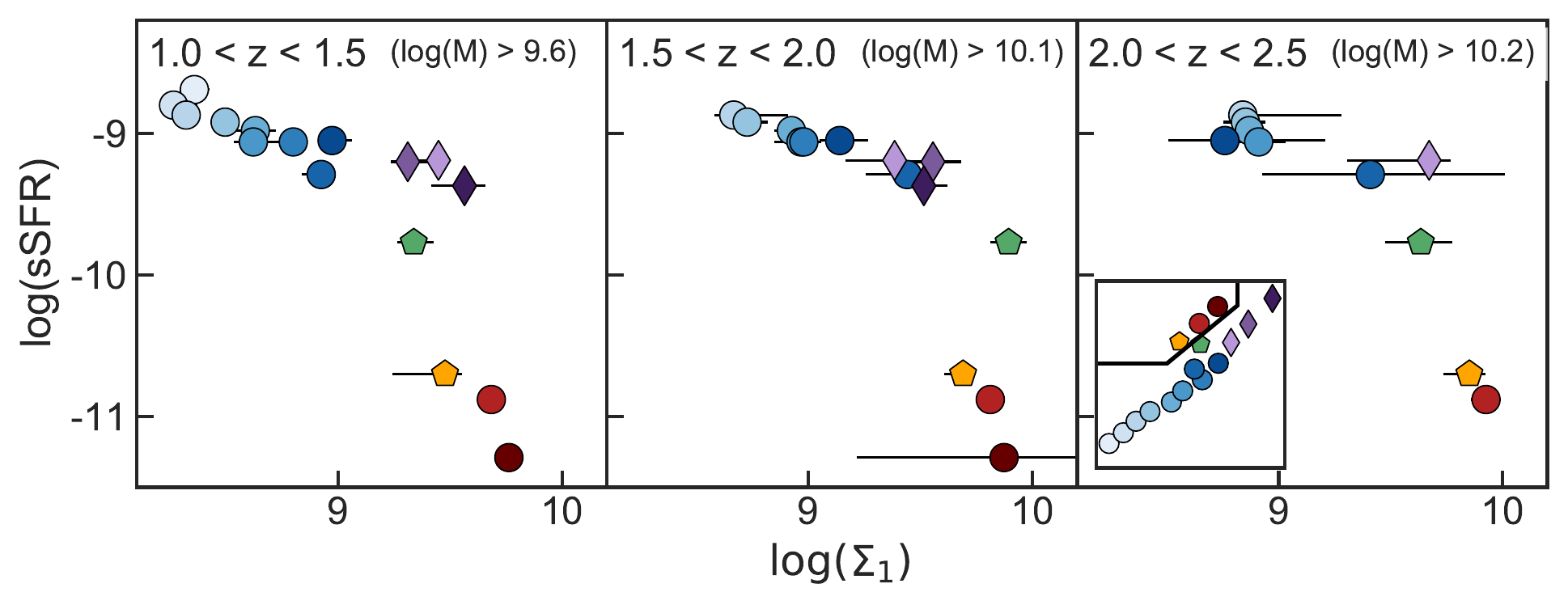}
    \caption{Median \sig as a function of sSFR for all galaxy groups. Each panel shows a different redshift interval (and has a corresponding mass completeness level). We see broad agreement with the \citet{barro17} ``L-shaped track": star-forming galaxies increase \sig as sSFR gradually decreases, and transitioning and quiescent galaxies have uniformly high \sig.}
    \label{fig:sig-sSFR}
\end{figure*}

In Figure~\ref{fig:sig-sSFR}, we consolidate the \sig-mass trends seen in Figure~\ref{fig:groupMassSig} by showing the sSFR of each group as a function of the median \sig value of galaxies in that group. 
Each panel shows a different redshift range, and has a corresponding mass completeness cut as indicated at the upper right of each panel. This mass completeness cut corresponds to the most stringent mass completeness cut of any individual group at that redshift. We see broad agreement with the ``L-shaped track" from \citet{barro17}: sSFR decreases only slightly as star-forming galaxies increase their central densities, then over a relatively narrow range in \sig the sSFRs of galaxies plummets towards quiescence. The bend in this diagram, where a small change in \sig corresponds to a large change in sSFR, consists of massive dusty star-forming galaxies, green valley galaxies, and post-starburst galaxies. These transitional galaxy populations have similar \sig values but dramatically different sSFRs: dusty star-forming galaxies have sSFRs consistent with the star-forming population, green valley galaxies have slightly but not fully suppressed sSFRs, and post-starburst galaxies have low sSFRs close to those of fully quiescent galaxies. We discuss the structures of these transitioning galaxies further in Section~\ref{sec:discuss_qui}. 

The \sig value at which galaxies transition from star-forming to quiescent appears to be a strong function of redshift in Figure~\ref{fig:sig-sSFR}. 
We caution that this is in large part due to the fact that we take a different mass cut in each redshift interval to retain a mass-complete sample. Because \sig is correlated with mass \citep[e.g.,][]{fang13,barro17,suess20} and our sample does not include lower-mass galaxies at high redshift, our median \sig values for all galaxy types are slightly higher at higher redshift. At fixed mass, the \sig quenching threshold is a much slower function of redshift than it appears in Figure~\ref{fig:sig-sSFR}: in \citet{suess20}, we show that \sig decreases by just  $\sim0.13$~dex over this redshift range \citep[see also, e.g.,][]{barro17,chen20,estrada20}.

\section{Implications for the star-forming sequence}
\label{sec:discuss_sf}
In this section, we discuss the interpretation and implications of our measurements for the star-forming structural relations. We only consider groups 1 - 9, which lie in the star-forming region of the $UVJ$ diagram and have $A_v<1.5$. Groups 10 - 12--- the most massive and dusty star-forming galaxies--- and group 13, the green valley galaxies, are discussed in Section~\ref{sec:discuss_qui}. 

\subsection{Individual star-forming groups follow steep size-mass relations}
\label{sec:parallelrelations}
Figure~\ref{fig:groupMassSig} shows that each group of star-forming galaxies traces out a different mass range of the star-forming \sig-mass relation. As sSFR decreases, total mass and \sig increase; this indicates that, on average, star-forming galaxies gradually increase both their total mass and the mass in their central kiloparsec as they grow. The slope and scatter of each group of galaxies is similar to the slope and scatter of the overall star-forming \sig-mass relation.

In contrast, Figure~\ref{fig:groupMassSize} shows that each individual star-forming group appears to have a relatively steep slope in size-mass space, more similar to the quiescent best-fit relation than the star-forming relation. 
Directly fitting a size-mass relation to each individual galaxy group is challenging due to the clear truncation of our sample at low masses \citep[for details, see e.g.][]{mantz19}. 
Instead, we test whether each group of star-forming galaxies is better fit by the overall star-forming size-mass relation, or by a relation with the steeper quiescent size-mass slope. We first re-normalize the quiescent size-mass relation, sliding it over in mass to match the observed mass range of each group of star-forming galaxies. We then compute the sum of the root mean square deviations between our observed data and each potential fit. We compare these two values and find that all star-forming groups at all redshifts--- other than groups 7 \& 8 at high redshift, which have too few galaxies to reliably fit--- are better fit by a shifted quiescent relation than by the star-forming size-mass relation. While the data does not demand that each star-forming group lies on a relation with a steep slope similar to the quiescent relation, it does {support} that conclusion.

It is unlikely that this result is driven purely by sample selection effects. The slope of each star-forming group is steep due to a lack of massive, compact galaxies. However, these massive, compact galaxies have high surface brightness and are thus relatively easy to detect: if they exist, they should be included in our sample. Furthermore, we do not observe any systematic change in observed axis ratio along each star-forming size-mass relation; this indicates that the trends we see are not due to orientation effects. We additionally verify that these trends are not caused purely by orientation effects by examining the sizes and masses of mock observations of simulated star-forming galaxies viewed from different orientations \citep{price17}. We find that orientation effects do not preferentially scatter galaxies in the direction of the steep size-mass relations we observe for each star-forming galaxy group. 

In light of these observations, we propose a new way to look at the star-forming size-mass relation. Instead of viewing it as a single monolithic relation that holds for all star-forming galaxies, we instead suggest that {\it the global star-forming size-mass relation is composed of many parallel relations, each of which has a relatively steep slope}. As star-forming galaxies evolve, their SED shapes change, their sSFRs decrease, their masses increase, and their sizes increase mildly in order to move the group up to the next parallel relation.  
This view of the star-forming size-mass relation is effectively an example of ``Simpson's paradox," where trends in aggregated data (e.g., the entire $UVJ$ star-forming population) differ substantially from trends in non-aggregated data (e.g., groups of star-forming galaxies with similar SED shapes). 

\begin{figure*}
    \centering
    \includegraphics[width=\textwidth]{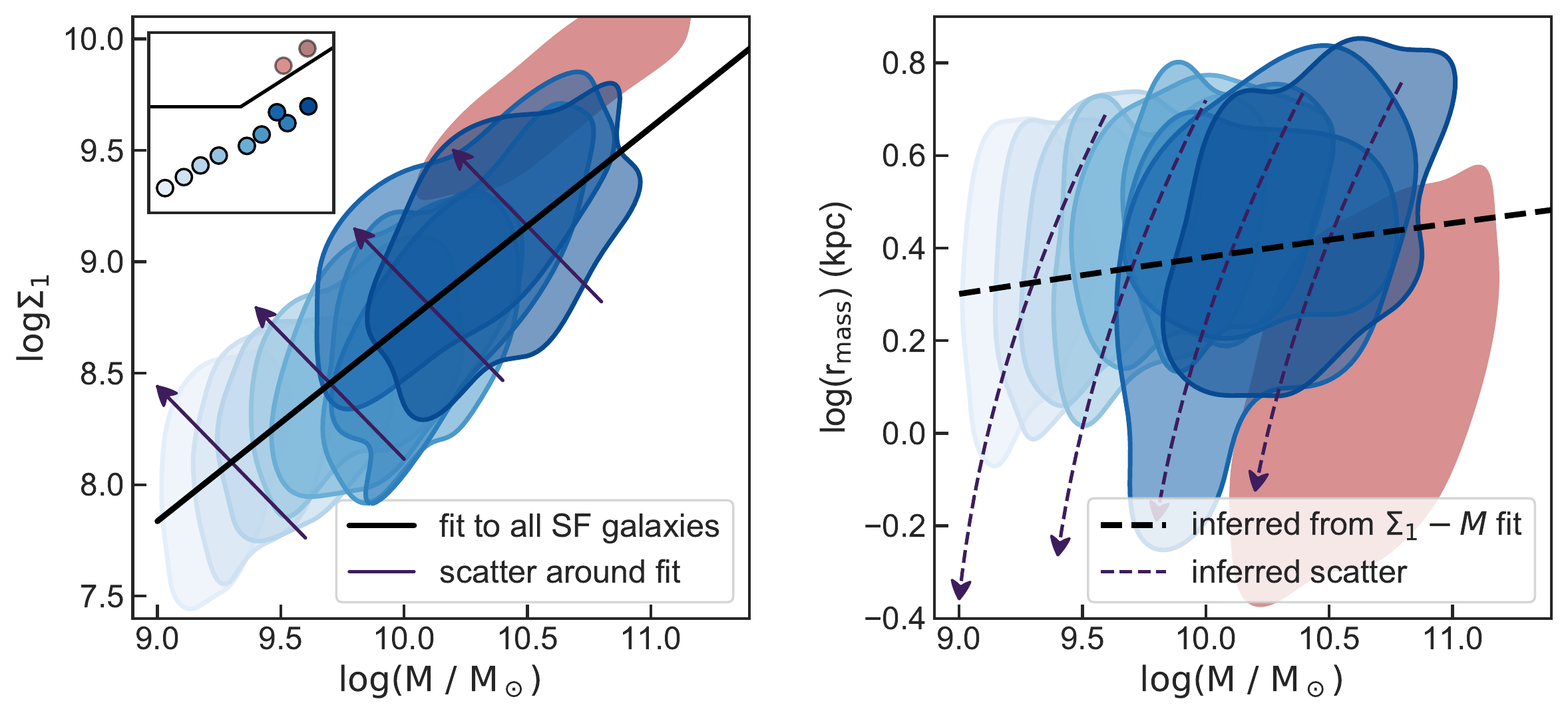}
    \caption{
    The distribution of star-forming galaxies in both the \sig-mass (left) and size-mass (right) planes. Blue shaded contours enclose the central 68\% of galaxies in each star-forming group (see inset $UVJ$); the red shaded contour shows the central 68\% of the two quiescent groups. Only $z<2$ galaxies with masses above the $z=2$ mass completeness limit are shown. 
    While each star-forming group traces out scatter around the \sig-mass relation, individual star-forming groups have steep size-mass slopes similar to the quiescent size-mass relation. This reveals that the star-forming size-mass relation is not a single monolithic relation: it is composed of {\it a series of many steep, overlapping parallel relations} for galaxies with different sSFRs.
    Furthermore, the distribution of star-forming galaxies in size-mass space can be fully described as a reflection of the \sig-mass relation and its scatter. 
    The thick solid line in the left panel shows a linear fit to the star-forming \sig-mass relation; the thin perpendicular lines show representative scatter around this relation. Dashed lines in the right panel show {\it this same fit \& scatter}, transformed to the size-mass plane by assuming that star-forming galaxies follow $n\sim1$ \sersic profiles. In both panels, arrows point towards more compact galaxies. This transformed \sig-mass fit \& scatter does an excellent job describing the data in the size-mass plane.
    }
    \label{fig:sfSlope}
\end{figure*}

This parallel relations picture also explains the correlation between galaxy size and sSFR at fixed mass \citep[e.g., Figure~\ref{fig:meanR_ssfr};][]{whitaker17}. Because each group has a relatively broad mass distribution that overlaps with neighboring groups, at fixed mass we are selecting galaxies from multiple different star-forming groups. However, because each group follows a steep size-mass relation, a fixed mass cut preferentially selects large galaxies from the higher-sSFR group (toward the ``upper end" of their parallel relation) and small galaxies from the lower-sSFR group (toward the ``lower end" of their parallel relation). This can be seen in right panel of Figure~\ref{fig:sfSlope}: a vertical mass cut selects the large galaxies in the ligher-blue, higher-sSFR groups, and small galaxies from the darker-blue, lower-sSFR groups.

\subsection{Explaining the size-mass relation as a reflection of the \sig-mass relation and its scatter}
We now turn to the connection between the size-mass and \sig-mass planes to understand the origin of the steep slopes we observe for each group of star-forming galaxies. \sig and half-mass radius for a given galaxy are connected by the shape and normalization of the galaxy's full mass profile. 
If we assume that star-forming galaxies follow $n=1$ \sersic profiles (e.g., exponential disks), there exists an exact mapping from one structural plane to the other. Because the \sig-mass relation has significantly less scatter than the size-mass relation \citep[e.g.,][]{fang13}, in what follows we use the best-fit \sig-mass relation and its scatter to predict how galaxies populate the size-mass plane. 

Equations~5 \& 6 of \citet{barro17} rewrite the \sersic profile as a relationship between total stellar mass and \sig. Rearranging these equations:
\begin{equation}
\label{eqn:guillermo}
\begin{split}
    \log\Sigma_1 &= \log \rm{M}_* - \log \pi +  \log\gamma(2n, b_n r_e^{-1/n}) \\
    & \approx \log \rm{M}_* - \log \pi + c_0 + c_1\log r_e + c_2 (\log r_e)^2
\end{split}    
\end{equation}
where the $c_i$ values are coefficients describing a second-order power-law fit to the incomplete gamma function.  
For an $n=1$ disk profile, \citet{barro17} find $c_0 = 0.31$, $c_1 = 0.96$, and $c_2 = 0.75$. 

Equation~\ref{eqn:guillermo} relates \sig to the stellar mass and effective radius.  
By assuming a best-fit \sig-mass relation of the form $\log\Sigma_1 = s \times \log\rm{M}_* + k$, we can substitute for \sig and re-write Equation~\ref{eqn:guillermo} as relationship between galaxy sizes and masses:

\begin{equation}
\label{eqn:sizemass}
    \log r_e = \frac{-c_1 + \sqrt{c_1^2 - 4 c_2 (c_0 + k + \log\pi + (s-1)\log\rm{M}_*} ) }{2c_2}
\end{equation}
Where we have disregarded the negative root of the quadratic because it produces unphysically small sizes ($\lesssim0.01$~kpc). We fit all star-forming galaxies in our sample (e.g., the blue points in Figure~\ref{fig:fullSample}) to find the coefficients $k$ and $s$ that describe the best-fit \sig-mass relation for all star-forming galaxies. We find:
\begin{equation}
\label{eqn:sigFit}
    \log{\Sigma_1} = (0.88 \pm 0.01) \times \log\rm{M_*} + (-0.11 \pm 0.10)
\end{equation}
These best-fit values are consistent with the \citet{barro17} star-forming \sig-mass relation within 1$\sigma$, despite differences in the methods used to calculate mass profiles and \sig values from multi-band imaging.

Next, we consider how {\it scatter} in the \sig-mass plane will transform to the size-mass plane. We note that this scatter is not dominated by measurement errors, but reflects true variations in the \sig values of galaxies at fixed mass. We assume that, at some given stellar mass $\rm{M}_* = \rm{M}_0$, the scatter around the \sig-mass relation can be described by a line perpendicular to the best-fit \sig-mass relation. Such a perpendicular line can be described by the equation:

\begin{equation}
\label{eqn:scatter}
    \begin{split}
        \log\Sigma_1 &= -\frac{1}{s} \times\log\rm{M_*} + \left( \frac{\rm{M}_0}{s} + \rm{M}_0 s + k  \right) \\
        &= s' \times \log\rm{M}_* + k'
    \end{split}
\end{equation}
By plugging these values of $s'$ and $k'$ into Equation~\ref{eqn:sizemass}, we can investigate how the scatter in \sig-mass space transforms to size-mass space.

We show this transformation from \sig-mass to size-mass space in Figure~\ref{fig:sfSlope}. The thick line in the left panel shows our best-fit \sig-mass relation; the thick dashed line in the right panel shows how this relation transforms to the size-mass plane using Equation~\ref{eqn:sizemass}. The thin lines in the left panel show representative scatter around the \sig-mass relation. These lines are spaced at equal $\log\rm{M}_*$ intervals and each span 0.3~dex in stellar mass; the arrows point towards more compact/smaller galaxies. The dashed lines in the right panel show how this representative scatter transforms to the size-mass plane using Equations~\ref{eqn:sizemass} \& \ref{eqn:scatter}; again, arrows point towards smaller/more compact galaxies.
Intriguingly, scatter in the \sig-mass plane does {\it not} translate to lines that are perpendicular to the overall size-mass relation. Instead, scatter in the \sig-mass plane corresponds to steep slopes in the size-mass plane. 

We then compare the predicted size-mass relation and scatter to our observations.
Each shaded blue contour in Figure~\ref{fig:sfSlope} shows the central 68\% of galaxies in a single star-forming group; the red contour shows the central 68\% of the two quiescent groups to guide the eye.
Figure~\ref{fig:sfSlope} includes only $1.0<z<2.0$ galaxies with masses above our mass completeness limit at $1.5<z<2.0$; this allows us to include high-sSFR galaxies, which typically have stellar masses below our $2.0<z<2.5$ completeness limit. We find that the size-mass relation predicted from the best-fit \sig-mass relation matches both the normalization and the shallow slope of the full star-forming sequence. While perhaps unsurprising--- we calculate both our \sig and half-mass radius data points from the same mass profiles--- this mapping between the size-mass and \sig-mass relations has not previously been demonstrated. More surprisingly, we find that the transformed scatter around the \sig-mass relation provides a good match to the steep size-mass slopes we observe for each group of star-forming galaxies. This allows us to provide a physical explanation for the ``parallel relations" view of the star-forming size-mass relation that we propose in Section~\ref{sec:parallelrelations}: star-forming galaxies with similar SED shapes populate steep tracks in size-mass space because those steep tracks trace out intrinsic scatter around the \sig-mass relation. 
This mapping also explains why the \sig-mass relation is tighter than the size-mass relation: as seen in Figure~\ref{fig:sfSlope} and Equation~\ref{eqn:guillermo}, the same $\delta\log\rm{M}_*$ produces a relatively small range in \sig but a large range in $\log r_e$ \citep[see also Equation 13 of][]{barro17}.

In Appendix~\ref{appendix:n_transform}, we show that our transformation from the \sig-mass plane to the size-mass plane does not strongly depend on our choice of \sersic index $n$ as long as it is within typical values for star-forming galaxies. While the slopes and normalizations differ slightly for $n=0.5$ and $n=2$ than for $n=1$, the predictions remain qualitatively similar and remain consistent with our data.

\subsection{The growth of star-forming galaxies}
Several previous studies have suggested that the tight scatter and mild redshift evolution of the star-forming \sig-mass relation implies that galaxies evolve {\it along} the \sig-mass relation \citep[e.g.,][]{tacchella15_science,barro17,woo19,chen20}. 
At the same time, other studies have concluded that star-forming galaxies evolve along the size-mass relation (\citealp[e.g.,][]{vandokkum15,lilly16,nelson19,wilman20}; see also \citealp{vanderwel09}). 
Figure~\ref{fig:sfSlope} allows us to unify these two pictures of the evolution of star-forming galaxies: galaxies grow along the size-mass relation because that growth directly corresponds to evolution along the \sig-mass relation. Changes in both the \sig-mass and size-mass relations are reflections of the same underlying changes in galaxy mass profiles.

We note that this size growth is relatively slow over the $1.0<z<2.5$ range we study in this paper: the change in star-forming half-mass radii at fixed mass is much smaller than expected from half-light radii \citep{suess19b}. This nearly self-similar growth is consistent with the roughly flat sSFR profiles that \citet{tacchella15_apj} and \citet{nelson16} observe in $\log{M_*/M_\odot}\lesssim10.5$ galaxies at these redshifts. 
Furthermore, this minimal size growth is consistent with our picture of star-forming galaxies evolving along the structural relations: the slope of the star-forming size-mass relation is only $\sim0.1$, implying that sizes increase only sightly as galaxies increase their stellar mass. Similarly, the slope of the \sig-mass relation is only slightly less than unity, implying that the mass in galaxy cores grows almost, but not quite, as quickly as total mass \citep[see also][]{woo19}. 

While star-forming sizes do not evolve rapidly over this redshift range, the strength of color gradients does \citep{suess19a,suess19b}. Part of this color gradient evolution could be caused by $A_{\rm{v}}$ gradients becoming stronger as galaxies become more massive \citep[e.g.,][]{whitaker17_dust}. Mild negative age gradients may also develop as galaxy sizes increase slightly. Probing the relative contribution of metallicity and age gradients could shed further light on how star-forming galaxies assemble their stellar mass. However, our current methods are unable to disentangle age, dust, and metallicity; spatially-resolved spectroscopy or rest-frame mid-IR imaging is required to break these degeneracies. {\it JWST} will enable the first such studies for a large sample of $z>1$ galaxies.

Finally, this view of star-forming structural evolution could explain the origin of the steep slope of the quiescent size-mass relation. If some massive star-forming galaxies quench their star formation without significantly altering their structures \citep[e.g., Section~\ref{sec:discuss_qui};][]{wu18}, then the steep size-mass slope of individual star-forming groups could be preserved as galaxies quench. Galaxies could thus join the red sequence with a steep size-mass slope {\it already in place}. 
Minor merger growth after quenching produces steep tracks in the size-mass plane \citep[e.g.,][]{bezanson09,naab09,vandokkum10,patel13}, further preserving and reinforcing the steep slope of the quiescent size-mass relation. 
Like for star-forming galaxies, this steep slope in size-mass space reflects scatter around the \sig-mass relation.

\section{Structures of Transitional Galaxies}
\label{sec:discuss_qui}

Next, we consider the structural properties of transitioning and quiescent galaxies. Our sample includes three distinct classes of possible quiescent progenitors: post-starburst galaxies, green valley galaxies, and dusty star-forming galaxies. All three of these transitional types lie in the region of Figure~\ref{fig:sig-sSFR} where a small change in \sig corresponds to a large change in sSFR.

\begin{figure*}
    \centering
    \includegraphics[width=\textwidth]{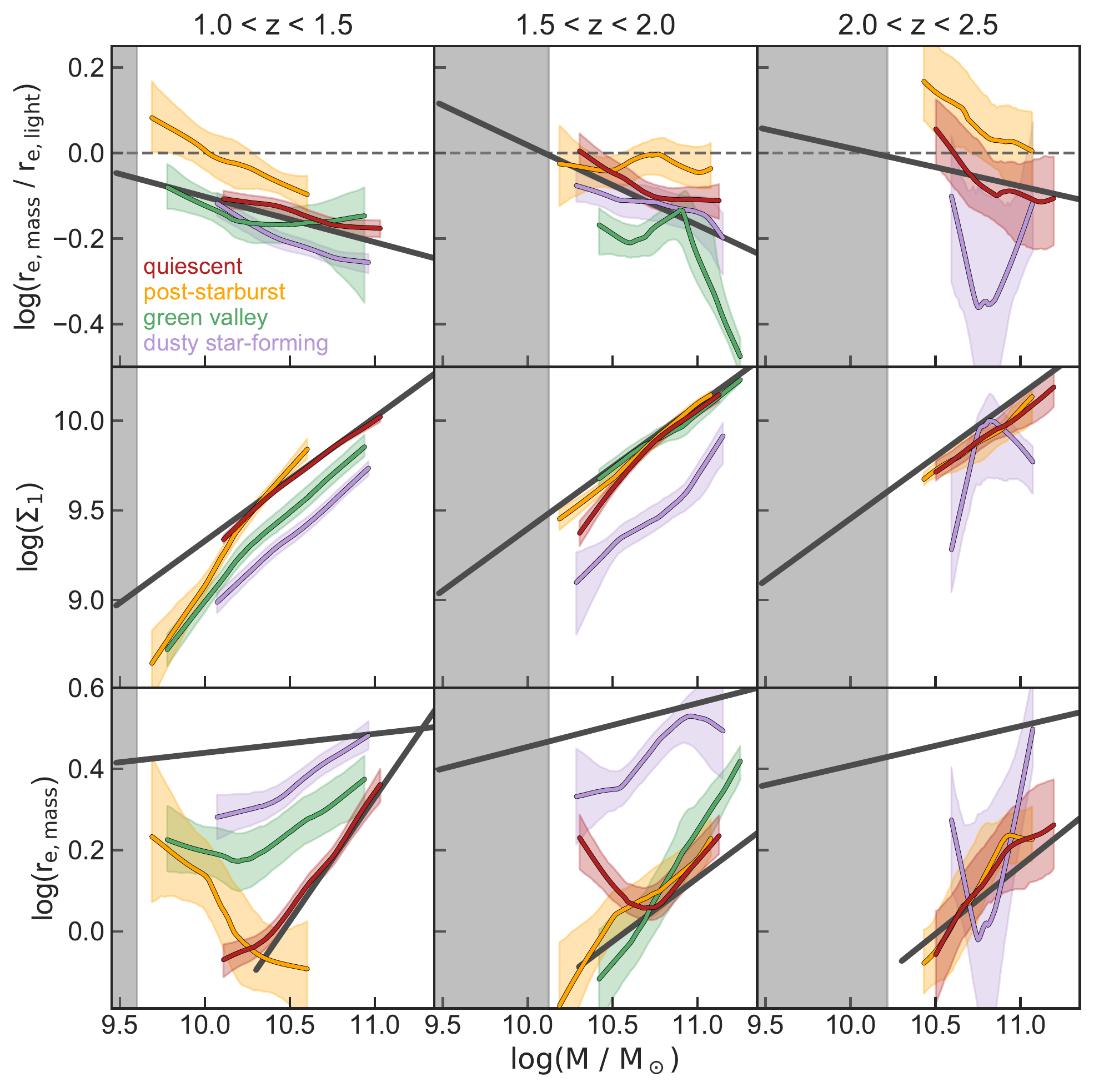}
    \caption{Color gradient strength (top row), central density (middle row) and half-mass radius (bottom row), for post-starburst galaxies (yellow), green valley galaxies (green), fully quiescent galaxies (red), and dusty star-forming galaxies (purple). Scatter points for each group of galaxies have been smoothed using LOWESS regression; the shaded regions represent the 1$\sigma$ confidence interval from 500 bootstrap realizations of the LOWESS smoothing. A given group is only shown if it contains $>10$ galaxies at that redshift; this excludes green valley galaxies at $2.0<z<2.5$. Grey lines show the best-fit color gradient-mass relations (Figure~\ref{fig:groupRatio}), the quiescent \sig-mass relation \citep{suess20}, and the star-forming and quiescent size-mass relations \citep{suess19a}. }
    \label{fig:transition}
\end{figure*}

Figure~\ref{fig:transition} shows the color gradient strengths (as measured by $r_{\rm{e,mass}}/r_{\rm{e,light}}$), \sig values, and half-mass radii of quiescent and transitional galaxies as a function of stellar mass. Quiescent galaxies (groups 15 and 16) are shown in red; post-starburst galaxies (group 14) are shown in yellow; green valley galaxies (group 13) are shown in green, and dusty star-forming galaxies (groups 10, 11, and 12) are shown in purple. We smooth the data points for each type of galaxies using locally weighted scatterplot smoothing (LOWESS), and use bootstrap resampling to calculate the 16-84\% confidence interval. We only show results for groups that contain $>10$ galaxies at a given redshift; this cut excludes green valley galaxies at $2.0<z<2.5$. 

Figure~\ref{fig:transition} shows that the structural properties of transitioning galaxies present a complex picture: different groups have distinct structural properties, and there is obvious dependence on both stellar mass and redshift. 
This is in contrast to $UVJ$-selected star-forming and quiescent galaxies, whose structures do not evolve significantly over the redshift range studied here \citep[][]{suess19a,suess19b}. The following subsections discuss the structures of each type of transitional galaxies in detail.

\subsection{Post-Starburst Galaxies}
Post-starburst galaxies have low sSFRs and strong Balmer breaks; their spectra indicate that they recently shut off a major burst of star formation \citep[e.g.,][]{dressler83,couch87,zabludoff96,leborgne06}. 
Figure~\ref{fig:transition} shows that $\log{M_*/M_\odot}\gtrsim10$ post-starburst galaxies have sizes and \sig values consistent with quiescent galaxies. However, \psbs have flat or even slightly positive color gradients, unlike the negative color gradients we observe in both star-forming and quiescent galaxies at this mass and redshift \citep{suess19a}. 
These measurements agree with our previous results in \citet{suess20}, which used a slightly different $UVJ$ color-based selection method to identify \psbs \citep[see also][]{maltby18}.   
In \citet{suess20}, we argue that these structural properties are consistent with post-starburst galaxies quenching via a rapid ``compaction"-like process, where elevated central star formation rates flatten existing negative color gradients, raise \sig, and decrease half-mass and half-light radii \citep[see, e.g.,][]{dekel14,zolotov15,tacchella16}.

In contrast to $z>1.5$, there are no high-mass ($\log{M_*/M_\odot}\gtrsim10.5$) \psbs in our sample at $z<1.5$. 
This result is consistent with previous studies which find that the number density of massive \psbs drops by a factor of $\sim2-5$ between $z\sim2$ and $z\sim1$ \citep[e.g.,][]{whitaker12_psb,wild16,belli19}. High-mass transitional galaxies do exist in our sample at low redshift; however, they are classified as either green valley or dusty star-forming galaxies, not \psbs.

At $1.0<z<1.5$, we see a significant population of low-mass ($\log{M_*/M_\odot}\lesssim10$) \psbs \citep[in agreement with number density measurements from][]{wild16}. These low-mass \psbs have distinct structures from their more massive counterparts: while massive \psbs have small sizes and high \sig values, low-mass \psbs have relatively larger sizes and lower \sig values.  
These results agree with \citet{maltby18}, who used light profiles to conclude that low-redshift \psbs tend to have low masses and large sizes. 
Our results additionally show that low-redshift, low-mass \psbs have more positive color gradients (e.g., relatively bluer centers) than their higher-mass counterparts.
The distinct structural properties of these low-mass \psbs may imply they are quenching via a different process than high-mass \psbs. In particular, environmental effects are typically more important for $M_* \lesssim 10^{10}M_\odot$ galaxies, and may be playing an additional role in transforming both the sSFRs and structures of these galaxies \citep[e.g.,][]{peng10,schawinski14,ji18}. Both \citet{socolovsky18} and \citet{moutard18} show that low-mass, low-redshift \psbs are more likely to reside in dense environments than in isolation, confirming that environmental effects are important for these galaxies.

\subsection{Green Valley Galaxies}
Green valley galaxies have intermediate star formation rates and lie along the diagonal edge of the $UVJ$ quiescent box; previous studies have suggested they are on their way towards quiescence \citep[see, e.g.,][]{patel11,fang18,gu18}. We note that the term ``green valley" was originally used to refer to transitional galaxies selected from color-magnitude or color-mass space \citep[e.g.,][]{bell04,martin07,wyder07,faber07,mendez11,goncalves12}. However, these single-color selections can include both dusty star-forming galaxies \citep[e.g.,][]{brammer09} and multiple different types of transitional galaxies \citep[e.g.,][]{schawinski14}. Later studies therefore introduced a second color, selecting green valley galaxies from color-color space \citep[e.g.,][]{patel11,arnouts13,fang18,gu18}. The green valley sample we discuss in this section is selected by rest-frame SED shape (Section~\ref{sec:sed_groups}), and is more similar to a color-color selected sample than a color-magnitude selected sample. 

Our composite SED fits show that the green valley galaxies in our sample have longer star formation timescales and larger $A_{\rm{v}}$ values than our post-starburst group. 
Like \psbs, we see significant changes in the green valley population across redshift. Unlike \psbs, however, we find that green valley galaxies are more abundant towards lower redshift.
There are just nine green valley galaxies in our sample at $z>2$, too few to show in Figure~\ref{fig:transition}. 

At $1.5<z<2.0$, green valley galaxies have high \sig values and relatively small sizes that are similar to both post-starburst and quiescent galaxies. However, green valley galaxies have stronger negative color gradients than either post-starburst or quiescent galaxies.
As discussed further in Section~\ref{sec:discuss_wrapup}, these differences in color gradient strength indicate that green valley and \psbs have different formation mechanisms.  
Furthermore, the fact that both green valley galaxies and star-forming galaxies have negative color gradients indicates that green valley galaxies do not have to significantly alter the radial distribution of their stellar populations in order to quench. We discuss the implications of these measurements for quenching in Section~\ref{sec:discuss_wrapup}.

Massive ($\log{M_*/M_\odot}\gtrsim10.5$) green valley galaxies at $1.0<z<1.5$ have slightly larger sizes and lower \sig values than their higher-redshift counterparts. The sizes and \sig values of these green valley galaxies are no longer consistent with the bulk of the quiescent population.
These differences are accentuated at low masses, where green valley galaxies are further offset towards large sizes and low \sig. At these low masses and redshifts, the sizes and \sig values of green valley galaxies are consistent with those of \psbs.
But again, green valley and \psbs have different color gradient strengths suggesting differences in the radial distributions of their stellar populations. Similar to low-mass \psbs, environmental effects are likely important for quenching low-mass green valley galaxies \citep[e.g.,][]{peng10}

We note that the large sizes of low-mass green valley and post-starburst galaxies appear to be responsible for the previously-observed flattening of the quiescent size-mass relation at $\log{M_*/M_\odot}\lesssim10$ \citep[e.g.,][]{cappellari13,vanderwel14,norris14,whitaker17}. If we used $UVJ$ colors to bin our sample into just two groups, star-forming or quiescent, most post-starburst galaxies and some green valley galaxies would be classified as quiescent. The large sizes of these transitional galaxies at low masses would drag the median inferred size of quiescent galaxies up.

\subsection{Dusty Star-Forming Galaxies}
The dusty star-forming galaxies in our sample have \av$>1.5$~mag, high stellar masses, and relatively high \sig values (Figure~\ref{fig:sig-sSFR}). We note that on average, the  dusty star-forming galaxies in our sample are less extreme than submillimeter-selected dusty star-forming samples, with $\rm{SFR}\sim200M_\odot\rm{yr}^{-1}$ and modest submillimeter fluxes \cite[e.g.,][]{spitler14}.
Figure~\ref{fig:transition} shows that the structures of dusty star-forming galaxies depend on redshift. At $z\lesssim2$, dusty star-forming galaxies tend to have mild negative color gradients, relatively large sizes, and \sig values below the quiescent population. At $z>2$, however, we find that most dusty star-forming galaxies have small sizes, high \sig values, and strong negative color gradients. 

These compact $z>2$ dusty star-forming galaxies have similar properties to ``blue nugget" or ``compact star-forming galaxy" samples that are selected by their small sizes or high \sig values \citep[e.g.,][]{barro14,vandokkum15}.
The compact structures, high stellar masses, and rapid number density evolution of these galaxies has led previous studies to conclude that compact star-forming galaxies are likely progenitors of compact quiescent galaxies \citep[e.g.,][]{barro14,nelson14,vandokkum15,barro17_alma,barro17}.
Interestingly, we have recovered a structurally similar sample despite not explicitly selecting for compactness.
This implies that compact structures are actually the norm for massive dusty star-forming galaxies at $z>2$. Moreover, we show that the mass profiles of these compact galaxies are even more concentrated than their light profiles.

The strong negative color gradients we observe in $z>2$ dusty star-forming galaxies are consistent with the idea that these galaxies are experiencing a strong dust-obscured central starburst before reaching quiescence \citep[e.g.,][]{barro17_alma}. IFU and ALMA studies of a small sample of dusty star-forming galaxies indicate that they have extremely dust-obscured centers and strong radial $A_{\rm{v}}$ gradients \citep[e.g.,][]{nelson16_dust,barro17_alma,tacchella18,tadaki20}. While a central starburst also creates positive age gradients, these strong $A_{\rm{v}}$ gradients likely dominate to create our observed negative color gradients.

At $1.0<z<2.0$, the dusty star-forming galaxies in our sample have mild negative color gradients, relatively large sizes, and lower \sig values.
These galaxies are consistent with an extension of the star-forming sequence discussed in Section~\ref{sec:discuss_sf}: dusty star-forming galaxies have slightly higher masses, lower sSFRs, smaller median sizes, and higher \sig values than their lower-\av counterparts. But unlike high-redshift dusty star-forming galaxies, low-redshift dusty star-forming galaxies do not have dramatically different structures than galaxies with slightly higher sSFRs.

The lack of compact dusty star-forming galaxies in our sample at $z<2$ is broadly consistent with \citet{barro13} and \citet{vandokkum15}, who found that the number density of compact star-forming galaxies decreases rapidly below $z\sim2$. We do, conversely, see some extended dusty star-forming galaxies at $z>2$. Like their lower-redshift counterparts, these galaxies have relatively lower \sig and more mild negative color gradients. From Figures~\ref{fig:groupMassSize} \& \ref{fig:groupMassSig}, we see that these extended galaxies are typically found in group 10, which has less dust and a lower sSFR than groups 11 \& 12. In Section~\ref{sec:discuss_wrapup}, we discuss the implications of both extended and compact dusty star-forming galaxies for quenching and the buildup of the red sequence.

Finally, we note that our size measurements for these dusty star-forming galaxies do not appear to be driven by observational biases.
First, inclination bias does not seem to play an important role: the observed axis ratios of the compact and extended $z>2$ dusty star-forming galaxies do not differ significantly \citep[though see][]{mowla19}. Second, our methods appear able to accurately recover the half-mass radii of dusty star-forming galaxies despite their highly dust-obscured centers. 
We test this by cross-matching our catalog to the rest-frame far-IR sizes of massive star-forming galaxies as measured by \citet{tadaki20}; these ALMA measurements are able to peer through dust and provide an independent and potentially less-biased view of the mass profiles of these galaxies. We find a median offset of just 0.15~dex between our half-mass radii and the \citet{tadaki20} rest-frame far-IR sizes, indicating that our methodology is not drastically overestimating galaxy sizes due to underestimated $A_{\rm{v}}$ gradients. We note that the half-light radii of galaxies in our sample are offset from the \citet{tadaki20} measurements by 0.4~dex.

\section{Implications for galaxy growth and quenching}
\label{sec:discuss_wrapup}

A growing number of recent studies have suggested the existence of two distinct pathways to quench galaxies, a ``fast" pathway in which galaxies stop their star formation abruptly, and a ``slow" path in which quenching happens more gradually \citep[e.g.,][]{barro13,barro14,schawinski14,wild16,carnall18,forrest18,wu18,rowlands18,woo19,belli19}. Fast-quenching galaxies tend to have post-starburst SEDs, while slow-quenching galaxies typically fall into the ``green valley" region between the quiescent and star-forming regions of the $UVJ$ diagram. In addition to differences in timescales, these studies have found that slow-quenching green valley galaxies are on average larger and diskier than fast-quenching post-starburst galaxies.

The above studies have ascribed these differences in structures and quenching timescales to two distinct physical quenching mechanisms operating in green valley and post-starburst galaxies. However, other studies suggest that we may see rapidly-quenching galaxies at high redshift because there is some intrinsic spread in galaxy star formation timescales, and at high redshift the universe is simply too young to observe any slow-quenching galaxies \citep[e.g.,][]{lilly16,abramson16}. 
In this case, the larger sizes of green valley galaxies can be explained through progenitor bias: star-forming galaxies are smaller at high redshift, so galaxies which quench rapidly at high redshift are smaller than galaxies which quench slowly at low redshift. 

In this section, we explore what our structural measurements imply for galaxy growth and quenching, and evaluate whether our observations support two distinct physical mechanisms for quenching. Our study adds multiple new pieces of information to this discussion. First, our sizes and \sig values are calculated from mass profiles, mitigating biases introduced by variations in color gradient strength between different galaxy types (Figure~\ref{fig:groupRatio}). Second, we are able to use the strengths of color gradients themselves as an additional tool to understand how galaxies grow and evolve. Third, our sample includes galaxies of all types across both the star-forming and quiescent sequences, instead of pre-selecting just a few types of transitional galaxies. In particular, this allows us to understand how massive dusty star-forming galaxies fit into the story of how galaxies quench.

Table~\ref{tab:summary_table} summarizes our structural measurements, discussed in detail in Sections~\ref{sec:discuss_sf} \& \ref{sec:discuss_qui}. Further details on the structures of post-starburst and quiescent galaxies can also be found in \citet{suess20}.

\begin{table}[]
    \centering
    \caption{Summary of our structural measurements, discussed in detail in Sections~\ref{sec:discuss_sf} \& \ref{sec:discuss_qui}.}
    \label{tab:summary_table}
    \includegraphics[width=.48\textwidth]{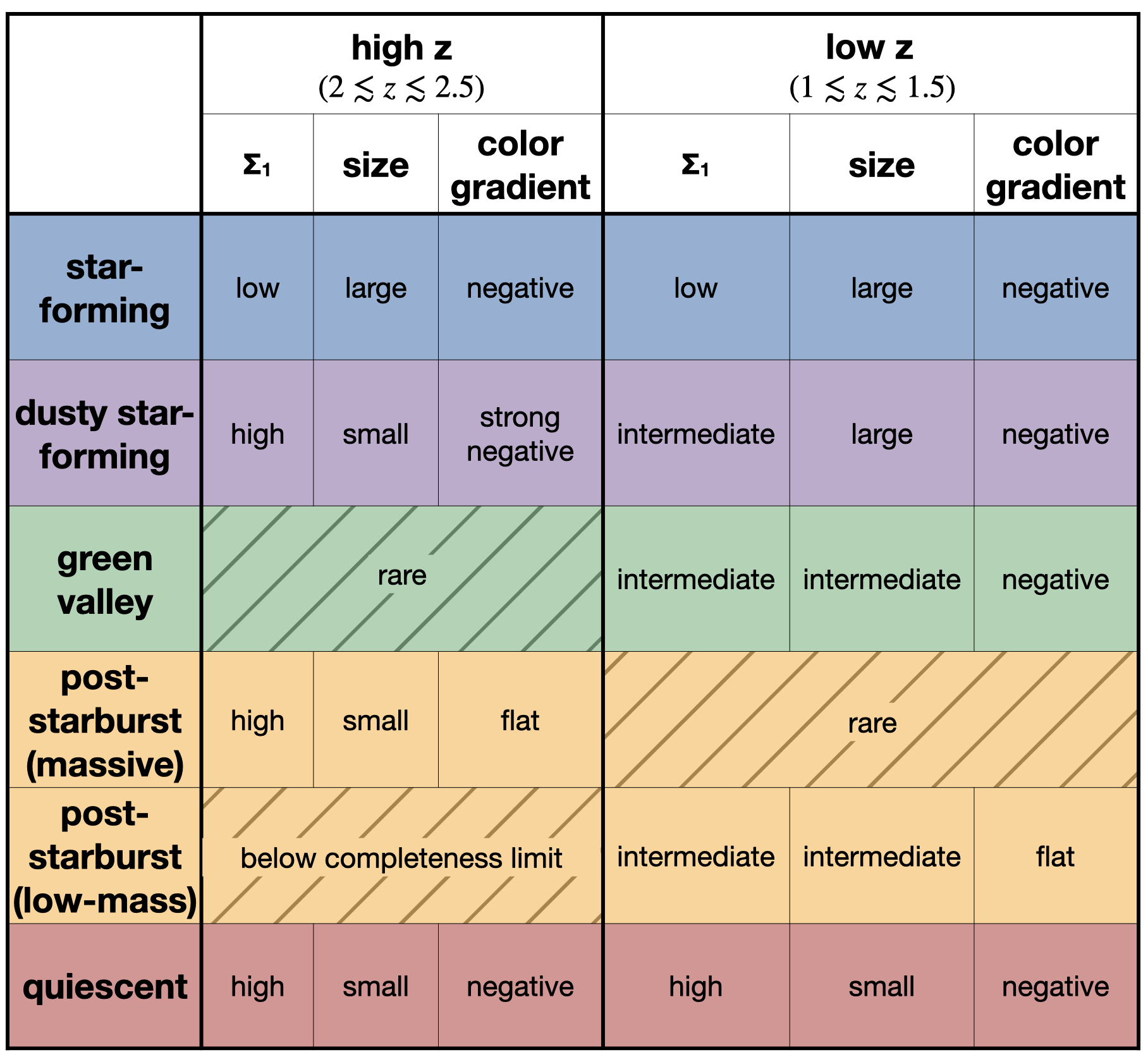}
    \tablecomments{For brevity we have only included our highest- and lowest-redshift intervals. At intermediate redshifts, both post-starburst and green valley galaxies have similar sizes as quiescent galaxies.}
\end{table}

\subsection{Two distinct quenching pathways} 
Here we assess whether our observations support multiple distinct physical mechanisms for quenching. If the physics responsible for quenching green valley and \psbs is the same--- \psbs just have rapid star formation timescales, whereas green valley galaxies have slower star formation timescales--- then any difference in their structures would be due to progenitor bias effects. In this scenario, green valley and \psbs should have comparable structures at fixed mass and redshift, because they are quenching from the same population of star-forming galaxies. 
At $1.5<z<2.0$ our sample includes both post-starburst and green valley galaxies with $\log{M_*/M_\odot}\gtrsim10$, where environmental effects are likely not dominant. This overlapping sample allows us to test this prediction. 

Figure~\ref{fig:transition} shows that the sizes and \sig values of post-starburst and green valley galaxies at $1.5<z<2.0$ are indeed consistent. However, the two classes of galaxies have {\it systematically different color gradients}. Because these color gradients describe radial variations in stellar population properties, these systematic color gradient differences indicate that post-starburst and green valley galaxies likely assembled their stellar mass via different physical processes. 
The differences between these slow- and fast-quenching galaxies {cannot} simply be ascribed to differences in their star formation timescales or overall dust content: they must have different formation mechanisms \citep[see also, e.g.,][]{barro13,woo19,belli19,wu20}. Our novel color gradient measurements thus support the idea that green valley and post-starburst galaxies represent two truly different pathways to quenching. 

\begin{figure*}
    \centering
    \includegraphics[width=.7\textwidth]{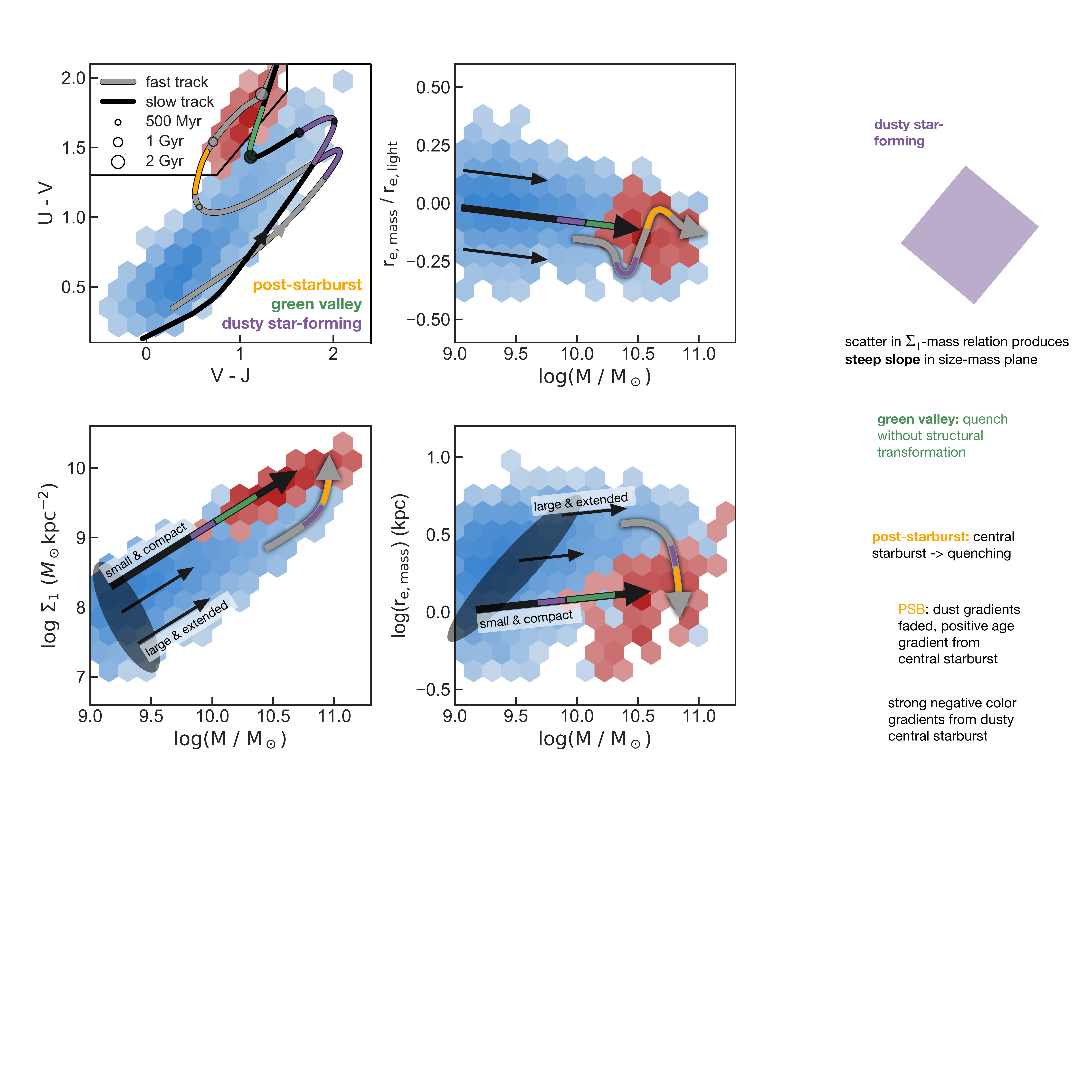}
    \caption{Schematic of our proposed growth and quenching pathways for the evolution of galaxies in the $UVJ$, \sig-mass, and size-mass planes at $z\sim1.5$. Red and blue shaded hexagons show the galaxies in our sample, binned into quiescent and star-forming groups using the \citet{whitaker12} $UVJ$ classification. Toy model $UVJ$ tracks show the evolution of a fast-quenching/post-starburst galaxy with $\tau=100$~Myr (grey) and a slow-quenching/green valley galaxy with $\tau=500$~Myr (black). {Shaded circles along the $UVJ$ tracks indicate 500 Myr, 1 Gyr, and 2 Gyr into the track's evolution. }
    The right two panels show how our structural measurements map onto these growth and quenching pathways. We propose that, on average, galaxies are born with some intrinsic scatter around the \sig-mass relation (corresponding to steep tracks in the size-mass plane), then grow parallel to the best-fit \sig-mass and size-mass relations. Green valley galaxies quench after this growth eventually pushes them across the high \sig quenching threshold; post-starburst galaxies quench after a compaction event causes their \sig values to increase and half-mass radii to decrease. The complex color gradient evolution we propose for post-starburst galaxies reflects a highly dust-obscured central starburst: once the burst stops and the strong $A_{\rm{v}}$ gradients fade, the products of this rapid quenching process should show flat or positive age gradients. The two pathways are shown operating at different masses purely for readability; both slow and fast quenching can produce quiescent galaxies across a wide stellar mass range. Again, we stress that these growth and quenching pathways represent the average evolution of galaxies in these structural planes--- the evolution of individual galaxies is likely much more complex and stochastic.}
    \label{fig:wrapup}
\end{figure*}

Figure~\ref{fig:wrapup} provides an illustration of how these fast and slow quenching pathways map onto the \sig-mass, size-mass, color gradient-mass, and $UVJ$ diagrams we discuss in this paper. 
Again, we emphasize that these two pathways show proposed {\it average} growth and quenching mechanisms: the tracks that individual galaxies follow in these structural planes are likely more complex and stochastic. This figure also illustrates $z\gtrsim1.5$, where green valley and \psbs have similarly compact sizes; the relatively larger sizes of low-redshift green valley galaxies are discussed further in Section~\ref{sec:implications_buildup}. 

We create two toy models to illustrate these fast and slow quenching pathways in $UVJ$ space. 
These toy models assume a delayed-$\tau$ star formation history and use FSPS \citep{conroy09,conroy10} to generate model $UVJ$ colors. Following \citet{barro14} and \citet{belli19}, we assume that dust evolves following the star formation rate, A$_v$ peaks at 2.5~mag, and quiescent galaxies have A$_v=0.4$~mag. The black track shows a slow pathway, with $\tau=500$~Myr; the grey track shows a fast pathway, with $\tau=100$~Myr. The fast pathway goes through the post-starburst region of $UVJ$ space after just $\sim600$~Myr. The slow pathway joins the quiescent sequence through the green valley after $\sim2$~Gyr without crossing through the post-starburst region. Both of the fast and slow tracks cross through the dusty star-forming region; however, the fast track does so just $150$~Myr into its evolution, whereas the slow track does so after nearly a full gigayear. 

We note that altering our assumptions about dust evolution will change the exact locations of these tracks in $UVJ$ space. A lower maximum \av will cause the tracks to quench without going through the dusty star-forming region. Alternately, additional dust can cause the fast track to go close to the diagonal $UVJ$ quiescent line, near the green valley region. This implies that selecting green valley galaxies by their location in $UVJ$ space alone may include some dusty fast-quenching galaxies at high redshift \citep[see, e.g.,][]{zick18}. We emphasize again that our selection in this paper is based on SED shape, not $UVJ$ colors alone.

\subsection{Slow growth \& quenching}
\label{sec:slow_quenching}
In Section~\ref{sec:discuss_sf}, we discuss a slow average growth mode where star-forming galaxies gradually increase their masses, sizes, and \sig values as they form stars. The bottom panels of Figure~\ref{fig:wrapup} show how this growth maps onto the \sig-mass and size-mass planes, and the top right panel shows the corresponding color gradient-mass evolution. We suggest that galaxies are born at low stellar masses and low \sig values with some intrinsic scatter around the \sig-mass relation. This scatter around the \sig-mass relation maps onto the size-mass plane as relatively steep slopes similar to that of the traditional quiescent size-mass relation (Section~\ref{sec:discuss_sf}). This scatter may be caused by variations in the concentration of dark matter halos \citep[e.g.,][]{jiang19,chen20}.

Figure~\ref{fig:sfSlope} shows that as we go up in mass and down in sSFR, star-forming galaxies with similar SED shapes march up the \sig-mass and size-mass structural relations, roughly preserving the scatter around the \sig-mass relation and its reflection in size-mass space. Physically, this means that larger galaxies will stay larger and smaller galaxies will stay smaller as they form stars (as shown by small black arrows in Figure~\ref{fig:wrapup}). 
This small size increase with increasing mass implies that star-forming galaxies at these masses grow their outskirts slightly faster than their cores, creating shallow negative age gradients observable as negative color gradients. 

If this slow growth continues, galaxies will eventually evolve far enough along the \sig-mass, size-mass, and color gradient-mass relations to reach the high \sig values required for quiescence. Before becoming fully quiescent, these slow-quenching galaxies will pass through a transitional phase where they have intermediate sSFRs and \sig values. The structural measurements we present in Section~\ref{sec:discuss_qui} indicate that green valley galaxies represent this slow-quenching transitional phase.
Furthermore, at $z\lesssim2$ we find that dusty star-forming galaxies have relatively large sizes and low \sig values in between those of green valley galaxies and the rest of the star-forming sequence. 
These results suggest that some slow-quenching galaxies go through a dusty star-forming phase before passing through the green valley to the quiescent sequence. This agrees with the slow-quenching toy model shown in Figure~\ref{fig:wrapup}, which passes through the dusty star-forming region before reaching the green valley.

\subsection{Fast quenching}
\label{sec:fast_quenching}
The fast quenching pathway is associated with ``compaction"-type events \citep[e.g.,][]{dekel14,zolotov15,tacchella16}. In this picture, strong gas inflows or gas-rich major mergers funnel gas to the center of the galaxy, causing a central starburst. This centrally-concentrated starburst will increase \sig and decrease size, effectively lifting galaxies up to the quiescent \sig-mass relation \citep[e.g.,][]{woo15,woo19}. During a compaction event, we expect galaxies to show high \sig, small sizes, and negative color gradients due to their highly dust-obscured central starbursts. These structural properties are consistent with our observations of high-redshift compact dusty star forming galaxies (Section~\ref{sec:discuss_qui}).
After star formation stops and $A_{\rm{v}}$ gradients fade, this scenario predicts small sizes and high \sig values. The color gradients of these galaxies should also be less extreme: young stars created in the central starburst are no longer highly dust-obscured, resulting in a positive age gradient. These predicted sizes, \sig values, and color gradients are consistent with our observations of \psbs in Figure~\ref{fig:transition} \citep[though whether age gradients in post-starburst galaxies are flat or positive is observationally unclear, see e.g.,][]{setton20,deugenio20}. Thus, our measurements lead us to speculate that fast-quenching galaxies may go through a compact dusty star-forming phase then a post-starburst phase before ultimately joining the quiescent population. This matches the fast-quenching toy model shown in Figure~\ref{fig:wrapup}, which goes through the dusty star-forming region of $UVJ$ space before the post-starburst phase. Further studies directly comparing the number density and star formation histories of dusty star-forming galaxies and \psbs could directly test this theory, but are beyond the scope of our current paper.

Finally, we note that in Figure~\ref{fig:wrapup} we show this compaction process taking place at $\log{M_*/M_\odot}\sim10.75$. This is mostly for illustrative purposes: in principal, fast quenching could create quiescent galaxies over a wide mass range. However, this compaction process {\it is} expected to operate primarily at higher redshift, towards the peak of the cosmic star formation rate \citep[e.g.,][]{dekel09,keres09,zolotov15}.

\subsection{Implications for the buildup of the red sequence}
\label{sec:implications_buildup}
We have described two distinct routes for galaxies to join the red sequence: fast quenching through the post-starburst phase, and slow quenching through the green valley. Here, we describe how these two pathways work in tandem to build up the massive ($\log{\rm{M}_*/\rm{M}_\odot}\gtrsim10$) range of the red sequence.

At $z>2$, we find that transitional galaxies tend to belong to the post-starburst group, while green valley galaxies are much less abundant. This indicates that massive quiescent galaxies are produced primarily through the rapid quenching pathway \citep[in agreement with, e.g.,][]{wild16,belli19}. Because \psbs have small sizes and high \sig, most galaxies that join the quiescent sequence at $z>2$ will be compact. 
The dominance of the rapid pathway at high redshift is unsurprising given that the universe is still young: most slow-quenching galaxies have not yet had sufficient time to reach high \sig. 
However, we do begin to see some relatively large, low \sig dusty star-forming galaxies at this redshift. These galaxies, which in Section~\ref{sec:slow_quenching} we argue are part of the slow quenching pathway, point towards the emergence of a larger population of slow-quenching green valley galaxies towards lower redshift. 

Indeed, at $1.5 < z< 2.0$ we find that both green valley and \psbs contribute to the growth of the quiescent population. Both transitional types have small sizes and high \sig values at this redshift; again, this means that relatively compact galaxies are being added to the red sequence. Because the sizes and \sig values of newly-quenched galaxies do not evolve significantly from $z\sim2.5$ to $z\sim1.5$, progenitor bias does not contribute significantly to evolution of the quiescent sequence at these redshifts \citep[see also][]{suess19b,suess20}. At these intermediate redshifts, compact dusty star-forming galaxies have essentially disappeared from our sample \citep[in agreement with, e.g.,][]{barro14,barro16,vandokkum15}. Because these compact dusty star-forming galaxies galaxies are likely linked to the fast quenching pathway, their disappearance signals a lack of massive post-starburst galaxies in our sample at $z<1.5$.

At $1.0<z<1.5$, we find that massive quiescent galaxies are primarily produced via the slow pathway: green valley galaxies are common, whereas we do not find any massive post-starburst galaxies \citep[in agreement with, e.g.,][]{whitaker12_psb,wild16,rowlands18,belli19}. 
The lack of fast-quenching galaxies at low redshift makes sense given that the compaction events thought to trigger rapid quenching rely on rapid gas inflow to create a central starburst \citep[e.g.,][]{dekel14,zolotov15,tacchella16}. However, because the gas content of galaxies declines rapidly towards lower redshift \citep[e.g., ][and references therein]{tacconi20}, compaction events and rapidly-quenched galaxies should become relatively rare at lower redshift. 

Interestingly, we find that green valley galaxies have larger sizes and lower \sig values at $1.0<z<1.5$ than at $z>1.5$, indicating that more extended galaxies are being added to the quiescent population. The larger sizes of these green valley galaxies are not primarily caused by an increase in the sizes of their star-forming progenitors: the half-mass radii of star-forming galaxies do not grow significantly between $z\sim2.5$ and $z\sim1$ \citep{suess19a,suess19b}. Instead, as illustrated in Figure~\ref{fig:wrapup}, we expect that galaxies which are born above the median \sig-mass relation are the first to quench through the green valley--- these small, compact galaxies have a head start on reaching the high \sig values required for quiescence.  
As we move to lower redshift, there is more time for galaxies born below the median \sig-mass relation to grow to high \sig values. 
Adding these larger, lower \sig green valley galaxies to the quiescent population at $z<1.5$ will drive the median size of quiescent galaxies to increase, and the median \sig of quiescent galaxies to decrease \citep[e.g., progenitor bias,][]{vandokkum01,carollo13,lilly16}. Indeed, \citet{barro17} shows that the zeropoint of the quiescent \sig-mass relation decreases by $\sim0.2$~dex between $z\sim1.25$ and $z\sim0.25$. 
Because this progenitor bias effect is primarily caused by slow-quenching green valley galaxies, it is most important at low redshift. 
This explains the apparent discrepancy between \citet{suess20}--- who find that \sig does not depend on age for quiescent galaxies at $z\sim1.5$--- and \citet{tacchella17}, who find that \sig {\it does} depend on age at $z\sim0$. 

Finally, we note that the prevalence of slow quenching at $z\lesssim1$ naturally explains the occurrence of large quiescent disk galaxies at $z\lesssim0.5$ \citep[e.g.,][]{vandenbergh76,couch98,dressler99,masters10,bundy10}. Slow-quenching green valley galaxies appear to shut down their star formation without significantly altering their structures \citep[e.g., Section~\ref{sec:slow_quenching};][]{wu18}, allowing them to retain the disky structures of their star-forming progenitors as they quench. The cutouts in Figure~\ref{fig:images} show that many low-redshift green valley galaxies are indeed large and disky, resulting in large disky quiescent galaxies. 

In summary, we suggest that the $\log{\rm{M}_*/\rm{M}_\odot}\gtrsim10$ quiescent sequence is built up primarily by rapidly-quenching post-starburst galaxies at $z\gtrsim1.5$, then by slow-quenching green valley galaxies at $z\lesssim1.5$. Progenitor bias effects become more important at lower redshift as larger green valley galaxies begin joining the quiescent population. Compact dusty star-forming galaxies at high redshift are primarily associated with fast quenching, whereas extended dusty star-forming galaxies at low redshift seem to be more consistent with green valley progenitors. Finally, we note that our sample includes low mass ($\log{\rm{M}_*/\rm{M}_\odot}\lesssim10$) galaxies only at $1.0<z<1.5$. We find that low-mass transitional galaxies can be found in either the post-starburst or green valley groups, but tend to have large sizes and low \sig values. These galaxies may be produced via environmental quenching processes \citep[e.g.,][]{peng10,schawinski14,ji18}.

\section{Conclusions}

In this paper, we move beyond a bimodal view of the galaxy size-mass and \sig-mass relations to study the masses, half-mass radii, \sig values, and color gradient strengths of a wide range of galaxy types. 
We use a clustering algorithm to separate a sample of $\sim7,000$ galaxies at $1.0<z<2.5$ into sixteen groups with similar rest-frame SED shapes. Different groups of galaxies have distinct stellar masses, sSFRs, ages, and dust content. These groups represent different phases of galaxy evolution, ranging from young unobscured star-forming galaxies to old quiescent galaxies. We then use these groups to investigate how different types of galaxies populate the size-mass and \sig-mass structural planes. 

We find that the strength of radial color gradients differs for each group of galaxies. In \citet{suess19a}, we established that more massive galaxies have more strongly negative color gradients; the color gradients of most galaxy groups are fairly well-predicted by this color gradient - mass relation. However, we find two outliers: post-starburst galaxies have flatter color gradients than expected from their mass and redshift, while extremely dusty star-forming galaxies tend to have stronger negative radial color gradients than expected. The fact that the strength of color gradients varies between different groups of galaxies underscores the importance of using mass profiles and half-mass radii: using light profiles and half-light radii would result in biased structural measurements. 

We show that each group of galaxies populates a distinct and fairly localized region of size-mass space. Moving beyond a simple ``star-forming vs quiescent" bimodal sample selection reveals that, on average, galaxy sizes and \sig values vary smoothly with SED type. 
In particular, we find that the size of galaxies at fixed mass is a gradual function of sSFR: there is not a sudden jump from large star-forming galaxies to small quiescent galaxies, but instead a gradual decrease in average size towards lower sSFRs. 

We find that each group of star-forming galaxies appears to lie on a steep size-mass relation, more consistent with the quiescent size-mass slope than the shallow overall star-forming slope. These steep relations can be explained by considering the size-mass plane as a reflection of the \sig-mass plane. We show that, if we assume star-forming galaxies follow simple $n=1$ \sersic profiles, the best-fit \sig-mass relation predicts both the normalization and the shallow slope of the star-forming size-mass sequence. Scatter around the best-fit \sig-mass relation produces steep tracks in the size-mass plane that agree with our data for each star-forming group.
These results allow us to introduce a novel way to view the star-forming size-mass relation. Instead of a single monolithic relation which holds for all star-forming galaxies, we propose that the canonical star-forming size-mass relation is comprised of a series of steep, overlapping parallel relations. Each one of these parallel relations reflects intrinsic scatter around the star-forming \sig-mass relation; this scatter may be caused by variations in dark matter halo properties. These results support a growth mode where, on average, star-forming galaxies evolve {\it along} the \sig-mass and size-mass relations: as star-forming galaxies increase their masses and decrease their sSFRs, they also increase their \sig values and slightly increase their sizes \citep[see also][]{vandokkum15,tacchella15_apj,barro17,woo19,nelson19,chen20,wilman20}.

Using our mass profiles and novel galaxy classification method, we also confirm previous results suggesting that the mass density within 1~kpc, \sig, is a powerful predictor of quiescence. Our groups of galaxies follow a ``L-shaped track" in \sig-mass space \citep{barro17}: \sig increases gradually along the star-forming sequence, then sSFR drops once galaxies reach a critical quenching threshold in \sig. The bend in this diagram, where small changes in \sig correspond to large changes in sSFR, is populated by transitional galaxies, including the most massive dusty star-forming galaxies, green valley galaxies, and post-starburst galaxies.

Our measurements support the view that there are (at least) two distinct ways for galaxies to reach the high \sig values required for quiescence.  
In the ``slow" pathway, galaxies continue to gradually grow along the star-forming structural relations until they naturally reach high \sig values. The intermediate sizes, \sig values, color gradients, and sSFRs of green valley galaxies indicate that they are undergoing this slow quenching process. This slow quenching picture predicts that the most compact galaxies quench first: this explains both the larger sizes of low-redshift green valley galaxies and the increasing importance of progenitor bias effects towards $z\sim0$. 
We also find that $z\lesssim2$ dusty star-forming galaxies have extended structures and mild negative color gradients, indicating that they may be linked to this slow quenching pathway. The long timescales required for this slow growth--- $\gtrsim2$~Gyr from simple toy models--- explains why it is most prevalent at lower redshifts.

In contrast, we find that post-starburst galaxies tend to have small sizes, dense cores, and flat color gradients. These flat color gradients are inconsistent with the negative color gradients found in green valley galaxies, indicating that the two populations have different radial distributions of their stellar populations and represent different pathways to quenching. These flat color gradients also indicate that post-starburst galaxies likely experienced some event that created young stars at their centers before they shut down star formation.
As discussed in detail in \citet{suess20}, the structures of these galaxies are consistent with a ``fast" quenching process triggered by a compaction-type event. The compact structures and strong radial color gradients of high-redshift dusty star-forming galaxies indicate that these galaxies are also part of this fast quenching pathway, and may represent the ``burst" before galaxies are observed as post-starburst. Future work directly comparing the number densities and star formation histories of post-starburst and compact dusty star-forming galaxies can directly test this theory. Both massive post-starburst galaxies and compact dusty star-forming galaxies are absent from our sample at $z\lesssim1.5$, indicating that this rapid quenching pathway is primarily operational at high redshift.

This paper demonstrates the power of moving beyond simple ``star forming vs. quiescent" sample selections, measuring structural properties from mass profiles, and introducing radial color gradients as an additional probe of galaxy structural evolution. Using these novel approaches, we have gained significant insights into galaxy growth and quenching. The next step forward in understanding the structural evolution of galaxies is decomposing our observed color gradients into age, metallicity, and $A_{\rm{v}}$ gradients which can be more straightforwardly interpreted and compared to numerical predictions. Direct measurements of these stellar population gradients will soon be possible with the advent of rest-frame mid-IR photometric surveys with {\it JWST}.

\bigskip
\acknowledgements 
This paper took a long time to take shape; KAS has a correspondingly long list of folks to thank for helpful and insightful discussions about this work, including Pieter van Dokkum, Lamiya Mowla, Rachel Bezanson, Joel Leja, Sirio Belli, Kate Whitaker, Drew Newman, Sandro Tacchella, Jenny Greene, Tom Zick, Sandy Faber, David Koo, Arjen van der Wel, Phil Hopkins, Ellie Abrahams, Rebecca Barter, and Michael Medford. {We thank the anonymous referee whose
suggestions improved this paper.} KAS would also like to thank Alison Suess \& Matt Rigsby, whose home served as a safe writing haven during the pandemic.
This work is funded by grant AR-12847, provided by NASA though a grant from the Space Telescope Science Institute (STScI) and by NASA grant NNX14AR86G. This material is based upon work supported by the National Science Foundation Graduate Research Fellowship Program under grant No. DGE 1106400.

\appendix
\section{The transformation from \sig-mass to size-mass for different choices of \sersic indices}
\label{appendix:n_transform}
\begin{figure}[ht]
    \centering
    \includegraphics[width=.4\textwidth]{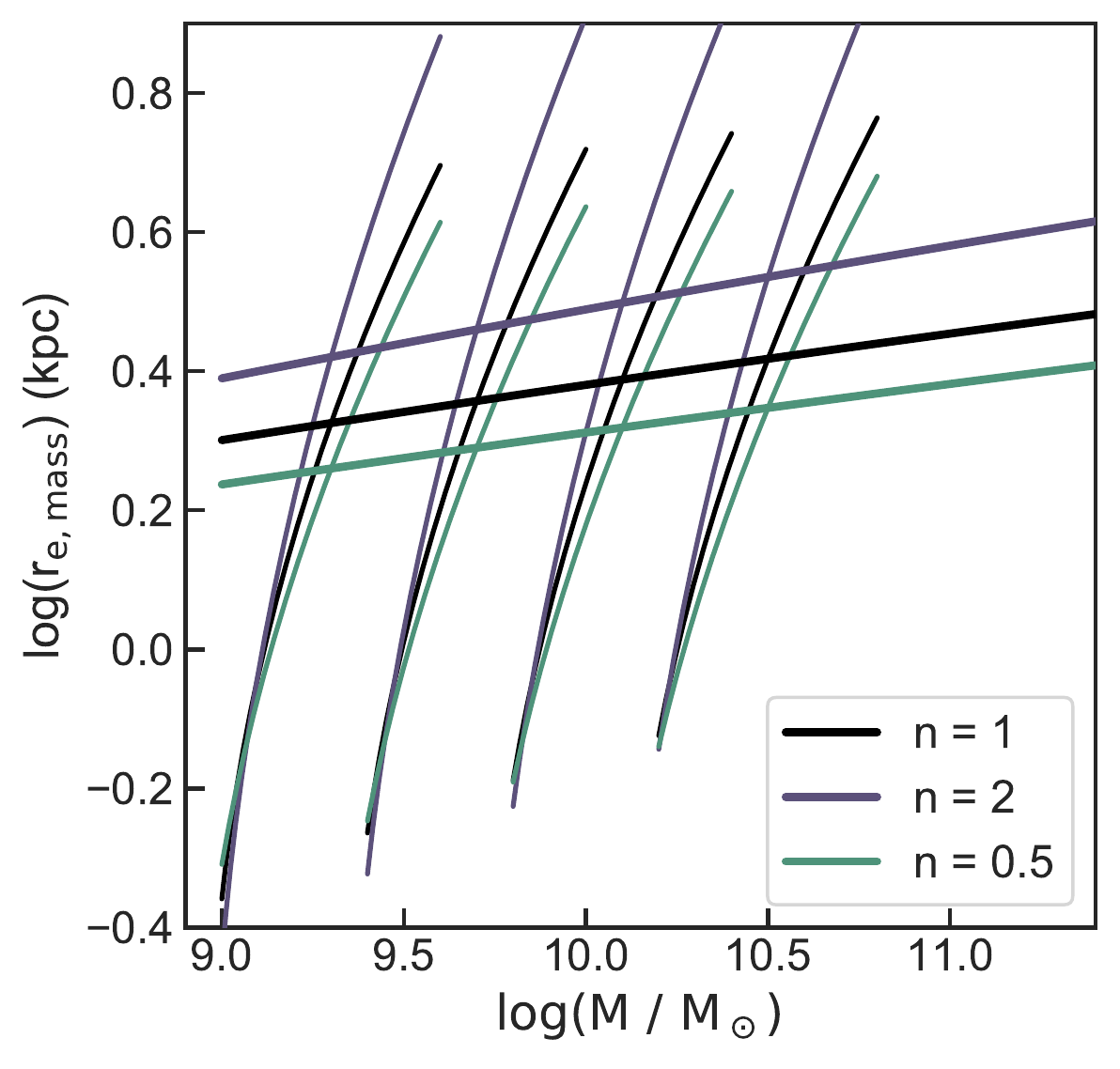}
    \caption{How the best-fit \sig-mass relation (Equation~\ref{eqn:sigFit}) transforms to the size-mass plane assuming different choices of \sersic index $n$. While the normalization of the best-fit relation changes slightly, the qualitative picture remains the same despite choice of $n$.}
    \label{fig:n_transform}
\end{figure}

Here, we show that the transformation from \sig-mass to size-mass space is relatively robust to our choice of \sersic index $n$, as long as $n$ is within typical values for star-forming galaxies. Figure~\ref{fig:n_transform} shows how both the best-fit \sig-mass relation and the scatter around that relation transform to the size-mass plane (e.g., Equations~\ref{eqn:sizemass} \& \ref{eqn:scatter}). The black $n=1$ lines are identical to those shown in the right panel of Figure~\ref{fig:sfSlope}. Changing the assumed \sersic index primarily affects the normalization of the best-fit size-mass relation. For $n=2$, the normalization is $\sim0.1$~dex higher; for $n=0.5$, the normalization is $\sim0.07$~dex lower; in both cases, the slope remains essentially unchanged. Changing $n$ also slightly affects the slope of the transformed scatter around the \sig-mass relation. For any choice of $n$, the qualitative picture remains the same as described in Section~\ref{sec:discuss_sf}.

\bibliographystyle{aasjournal}
\bibliography{massBib}

\end{document}